**Topical Review**

# Skyrmion-Electronics: Writing, Deleting, Reading and Processing Magnetic Skyrmions Toward Spintronic Applications

Xichao Zhang,[1] Yan Zhou,[1] Kyung Mee Song,[2] Tae-Eon Park,[2] Jing Xia,[1] Motohiko Ezawa,[3] Xiaoxi Liu,[4] Weisheng Zhao,[5] Guoping Zhao,[6] and Seonghoon Woo[7, †]

[1]*School of Science and Engineering, The Chinese University of Hong Kong, Shenzhen, Guangdong 518172, China*

[2]*Center for Spintronics, Korea Institute of Science and Technology (KIST), Seoul 02792, Korea*

[3]*Department of Applied Physics, University of Tokyo, Hongo 7-3-1, Tokyo 113-8656, Japan*

[4]*Department of Electrical and Computer Engineering, Shinshu University, Wakasato 4-17-1, Nagano 380-8553, Japan*

[5]*Fert Beijing Institute, BDBC, and School of Microelectronics, Beihang University, Beijing 100191, China*

[6]*College of Physics and Electronic Engineering, Sichuan Normal University, Chengdu 610101, China*

[7]*IBM Thomas J. Watson Research Center, Yorktown Heights, New York 10598, USA*

†E-mail: shwoo@ibm.com

(Dated: November 5, 2019)

## ABSTRACT

The field of magnetic skyrmions has been actively investigated across a wide range of topics during the last decades. In this topical review, we mainly review and discuss key results and findings in skyrmion research since the first experimental observation of magnetic skyrmions in 2009. We particularly focus on the theoretical, computational and experimental findings and advances that are directly relevant to the spintronic applications based on magnetic skyrmions, i.e. their writing, deleting, reading and processing driven by magnetic field, electric current and thermal energy. We then review several potential applications including information storage, logic computing gates and non-conventional devices such as neuromorphic computing devices. Finally, we discuss possible future research directions on magnetic skyrmions, which also cover rich topics on other topological textures such as antiskyrmions and bimerons in antiferromagnets and frustrated magnets.

(Some figures may appear in color only in the online journal)

# CONTENTS

# 1. Background

## 1.1. Topological Spin Textures

Topology is an important discipline of pure and applied mathematics, which also plays a significant role in the understanding of many real-world physical phenomena. In the field of magnetism and spintronics, the concept of topology is particularly important to the physics of some exotic magnetic spin textures (i.e., the so-called magnetic solitons) [1-7], including different types of magnetic skyrmions and vortex-like spin textures, where the behaviors of textures are determined or affected by their topological characteristics. These topologically non-trivial spin textures can be one-dimensional (1D), two-dimensional (2D) and three-dimensional (3D) in real space, and usually carry integer or half-integer topological charges determined by their spin textures in the topological space [1-22]. For most cases, the well-studied topological spin textures are those localized in 2D or quasi-2D magnetic thin films and multilayers [11, 19, 21, 23-36], because that magnetic thin films and multilayers are better understood and preferred for building nanoscale spintronic applications. Several recent works have also demonstrated promising properties of 3D topological spin textures [10, 37-50], leading to the prediction that 3D spin textures might be used in combination with 2D textures in the future. Indeed, the 1D magnetic solitons have also been investigated in the field [9, 12-17, 51, 52], which revealed many fundamental properties of topologically non-trivial objects. For example, the chiral helicoidal soliton lattice was experimentally observed in a chiral helimagnet in 2012 [52], which was theoretically envisioned by I. E. Dzyaloshinskii in 1964 [15]. In this review, however, we will focus on and limit our discussion to 2D topological spin textures in magnetic materials. Figure 1 shows various exemplary topological spin textures that can be found in such 2D or quasi-2D magnetic materials. Note that quasi-2D system means the variation of spin texture along the thickness direction can be ignored, such as in the thin film with a finite thickness.

The concept of skyrmion model was first proposed in the field of nuclear physics by T. H. R. Skyrme [53] in 1962. In 1989, Bogdanov and Yablonskiĭ for the first time predicted that chiral magnetic skyrmions can also be found in magnetic materials [16]. Then in 1994, Bogdanov and Hubert theoretically studied skyrmion structures in easy-axis magnetic materials with the Dzyaloshinskii-Moriya (DM) interaction [23, 54], where the DM interaction was found to be an important energy term to stabilize skyrmions, which will be discussed below. In 2001, Bogdanov and Rößler developed a phenomenological theory of chiral symmetry breaking in magnetic thin films and multilayers, and also predicted the existence of skyrmions in magnetic thin films and multilayers stabilized by induced DM interactions [18]. In 2006, Rößler and coworkers further theoretically demonstrated that skyrmion structures can be formed as spontaneous ground states in magnetic metals with DM interactions without the assistance of external fields or the proliferation of defects [20].

Due to the topological nature, spin textures including chiral magnetic skyrmions can be regarded as localized quasi-particles in magnetic materials and show topology-dependent static and dynamic properties [1-7, 55]. In the context of 2D and quasi-2D systems, the topological structure of a spin texture is generally characterized by the topological charge [55-57], which is also referred to as the Pontryagin number [2, 3]:

$$Q = \int d^2\boldsymbol{r}\, \rho(\boldsymbol{r}), \qquad (1)$$

where the topological charge density $\rho(\boldsymbol{r})$ reads

$$\rho(\boldsymbol{r}) = \frac{1}{4\pi} \boldsymbol{m}(\boldsymbol{r}) \cdot \left(\partial_x \boldsymbol{m}(\boldsymbol{r}) \times \partial_y \boldsymbol{m}(\boldsymbol{r})\right). \qquad (2)$$

Thus, the topological charge can be calculated based on the exact spin texture configuration according to

$$Q = \frac{1}{4\pi} \int d^2\boldsymbol{r} \cdot \boldsymbol{m}(\boldsymbol{r}) \cdot \left(\partial_x \boldsymbol{m}(\boldsymbol{r}) \times \partial_y \boldsymbol{m}(\boldsymbol{r})\right). \qquad (3)$$

The topological charge basically counts how many times the reduced local magnetization (i.e., the spin) $\boldsymbol{m}(\boldsymbol{r})$ wraps the 2D surface of a 3D ball in 3D space (i.e., the 2-sphere) as the coordinate $(x, y)$ spans the whole planar space. For example, as shown in Fig. 1, the Néel-type and Bloch-type skyrmions have topological charges of $Q = -1$ (for the core magnetization of $-m_z$), while the antiskyrmion of the same core magnetization has the opposite topological charge of $Q = +1$ (often referred as the antiparticle of skyrmions with $Q = -1$). It is also called as "*the skyrmion number*" [2, 55] for a series of skyrmion-like spin textures that will be discussed throughout this review.

In fact, chiral magnetic skyrmions are not limited to the case of topological charges $|Q| = 1$, and could be of any topological charge [24, 58-60]. For example, the skyrmionium can be regarded as a topological combination of a skyrmion with $Q = +1$ and a skyrmion with $Q = -1$, which carries a net topological charge of $Q = 0$. The skyrmionium structure was first studied in a theoretical work by Bogdanov and Hubert in 1999 [24]. It is also referred to as the target skyrmion [61]. The topological charge difference between the skyrmionium with $Q = 0$ and skyrmion with $Q = +1$ originates from their out-of-plane spin textures. A serious of theoretical works have suggested that skyrmioniums in magnetic materials can be manipulated by different external stimuli [11, 58, 62-73]. As other examples, the biskyrmion has a topological charge of $Q = -2$, which can be formed in some materials such as chiral bulk or frustrated magnets when two skyrmions with the same topological number ($Q = -1$ in this case) are approaching to each other. The meron and bimeron have topological charges of $Q = -0.5$ and $Q = -1$, respectively, where the bimeron consists of a meron with $Q = -0.5$ and an antimeron with $Q = -0.5$. The bimeron with $Q = -1$ can be regarded as a counterpart of skyrmions with $Q = -1$ in magnetic materials with easy-plane magnetic anisotropy, where magnetization prefer to lie in the plane [74]. We will discuss these skyrmion-like structures in detail in Sec. 5.2.

Obviously, these multifarious topological spin textures cannot be fully distinguished only using the topological charge $Q$, because there might be some degenerate states for a given value of the topological charge. Taking the skyrmion with $Q = -1$ as an example, there are additional degrees of freedom in the in-plane spin configuration. Therefore, one could calculate the vorticity number $Q_v$ and the helicity number $Q_h$, which completely characterize the topological spin texture accompanying the topological charge $Q$.

First, a point of the x-y space is parameterized as

$$x = r \cos \varphi, y = r \sin \varphi. \tag{4}$$

By applying the mapping $r = 0$ as $z \to -1$, $r = 1$ as $z = 0$, $r \to \infty$ as $z \to 1$, and $\lim_{r \to \infty} \boldsymbol{m}(x, y) = \lim_{z \to 1} \boldsymbol{m}(z, \phi)$ [56], we map the x-y space onto the 2D surface of a 3D ball parameterized by

$$x = \sqrt{1 - \cos^2 \theta} \cos \phi, y = \sqrt{1 - \cos^2 \theta} \sin \phi, z = z = \cos \theta, \tag{5}$$

with $z$ being defined as $r = (1 + z)/(1 - z) = (1 + \cos \theta)/(1 - \cos \theta)$. Therefore, we can re-write the local magnetization direction as

$$\boldsymbol{m}(\boldsymbol{r}) = \boldsymbol{m}(\theta, \phi) = (\sin \theta \cos \phi, \sin \theta \sin \phi, \cos \theta). \tag{6}$$

By substituting Eq. (6) into Eq. (3), we obtain the topological charge determined by $\theta$ and $\phi$, given as

$$Q = \frac{1}{4\pi} \int_\pi^0 \sin \theta \, d\theta \int_0^{2\pi} d\phi = -\frac{1}{4\pi} [\cos \theta]_\pi^0 [\phi]_0^{2\pi}. \tag{7}$$

From Eq. (7), it can be seen that the topological charge is actually determined by both the out-of-plane ($\theta$) and in-plane ($\phi$) spin textures. It should be noted that here we only consider the skyrmion solution where $\theta$ rotates $\pi$ when $r$ goes from zero to infinity. For the skyrmionium solution, $\theta$ rotates $2\pi$ when $r$ goes from zero to infinity [24] Hence, in order to describe the in-plane spin texture, we define the vorticity number as

$$Q_v = \frac{1}{2\pi} \oint_C d\phi = \frac{1}{2\pi} [\phi]_{\varphi=0}^{\varphi=2\pi}. \tag{8}$$

One also needs to introduce the helicity number, which is the phase appearing in

$$\phi = Q_v \varphi + Q_h. \tag{9}$$

Therefore, Eq. (6) can be as re-write as

$$\boldsymbol{m}(\theta, \varphi) = [\sin\theta\cos(Q_v\varphi + Q_h), \sin\theta\sin(Q_v\varphi + Q_h), \cos\theta]. \qquad (10)$$

Therefore, topological spin textures can be fully characterized by the three distinct numbers $(Q, Q_v, Q_h)$. Here it is worth mentioning that the spin structures of chiral spin textures can also be described and observed in reciprocal or momentum space [8, 75-77], which allows diffraction-based techniques using e.g. neutrons to be effectively used to characterize the crystal of magnetic textures [75, 78, 79]. For example, the hexagonal chiral magnetic skyrmion lattice in real space can be described in reciprocal space as a superposition of three helical modulations (meaning three basis vectors of the crystalline structure) [77], which can be observed as six-fold scattering patterns in reciprocal space [75, 78, 80].

Figure 2 depicts some examples of degenerate skyrmion textures with varying $Q_v$ and $Q_h$, where the basic skyrmion textures with the topological charge $|Q| = 1$ with different in-plane magnetization rotational senses are shown. For example, for the Néel-type skyrmion with a spin-up (i.e. pointing along the +z direction) and spin-down edge (i.e. pointing along the –z direction), its in-plane spins can point toward or away from the skyrmion core. For the Bloch-type skyrmion, its in-plane spins can form a circle in either a clockwise or counterclockwise fashion. However, for the antiskyrmion with $|Q| = 1$, although its helicity number can vary between 0 and $2\pi$, the spin texture actually has a two-fold rotational symmetry with respect to the skyrmion core. So, the antiskyrmion structure is effectively determined by its vorticity number [81]. It is noteworthy that skyrmions and antiskyrmions carry positive and negative $Q_v$, respectively. The topological spin textures with different topological structures build a large family and lead to the emerging field of *topological magnetism*, which promises new opportunities for magnetic and spintronic applications.

As the topological spin textures are non-collinear spin textures, their existence in magnetic materials is usually a result of delicate interplay among different energy terms. From the viewpoint of micromagnetism at zero temperature, the energy terms for common magnetic materials include the Heisenberg exchange interaction, dipolar interaction energy terms as well as magnetic anisotropy and Zeeman energy terms. The dominated Heisenberg exchange interaction favors the parallel or antiparallel alignment of the adjacent magnetic spins in most cases, so that it does not stabilize any non-collinear spin texture. However, the strong emergence of various other energy terms that may naturally exist in or be introduced to the magnetic materials can lead to the stabilization of non-collinear topological spin textures. In particular, the most important energy term, which has been studied for many years, is the asymmetric exchange interaction, i.e. the above mentioned DM interaction [82, 83]. The DM interaction favors a right angle between the adjacent magnetic spins, thus a delicate competition between the Heisenberg exchange and DM interactions could result in the formation of non-collinear domain wall structures and topological spin textures with a fixed rotation fashion (called "*chirality*"), such as the chiral magnetic skyrmion and magnetic helical state. It is noteworthy that the spin-helix length in magnetic materials with DM interactions is determined by $4\pi A/|D|$, where $A$ is the Heisenberg ferromagnetic exchange constant.

The homogeneous DM interaction can exist in bulk materials lacking inversion symmetry and is expressed as

$$E_{bDM} = d_{bDM} \sum_{\langle i,j \rangle} u_{ij} \cdot (\boldsymbol{m}_i \times \boldsymbol{m}_j), \qquad (11)$$

where $<i, j>$ denotes the nearest-neighbor sites, $\boldsymbol{m}_i$ and $\boldsymbol{m}_j$ are the reduced magnetic spin vectors at sites $i$ and $j$, respectively. $d_{bDM}$ is the bulk DM interaction coupling energy, $u_{ij}$ is the unit vector between $\boldsymbol{m}_i$ and $\boldsymbol{m}_j$. In the micromagnetic model [11], the bulk DM interaction reads

$$E_{bDM} = b \iint D_{bDM}[\boldsymbol{m} \cdot (\nabla \times \boldsymbol{m})]d^2\boldsymbol{r}, \qquad (12)$$

with $D_{bDM}$ being the continuous effective bulk DM interaction constant, and $b$ being the magnetic film thickness. The bulk DM interaction could result in the stabilization of Bloch-type skyrmions with ($Q = +1$, $Q_v = +1$, $Q_h = \pi$) and ($Q = -1$, $Q_v = +1$, $Q_h = 0$).

On the other hand, the DM interaction can also be induced at the interface between an ultrathin magnetic film and a non-magnetic film with a large spin-orbit coupling (SOC), given as

$$E_{iDM} = d_{iDM} \sum_{\langle i,j \rangle} (u_{ij} \times \hat{z}) \cdot (\boldsymbol{m}_i \times \boldsymbol{m}_j), \qquad (13)$$

where $d_{iDM}$ is the interface-induced DM interaction coupling energy, and $\hat{z}$ is the normal to the interface determined by Moriya's rule [83], oriented from the large SOC material to the magnetic film. In the micromagnetic model [11], the interface-induced DM interaction reads

$$E_{iDM} = b \iint D_{iDM}[m_z(\boldsymbol{m} \cdot \nabla) - (\nabla \cdot \boldsymbol{m})m_z]d^2\boldsymbol{r}, \qquad (14)$$

with $D_{iDM}$ being the continuous effective interface-induced DM interaction constant. The interface-induced DM interaction could result in the stabilization of Néel-type skyrmions with ($Q = +1$, $Q_v = +1$, $Q_h = \pi$) and ($Q = -1$, $Q_v = +1$, $Q_h = 0$).

Note that the structures of skyrmions stabilized by DM interactions are not only limited to Bloch- and Néel-types, as recent reports have reported the presence of intermediate type of skyrmions [84-87] stabilized by other DM interactions. Also, the Dresselhaus SOC in materials with bulk inversion asymmetry can generate a type of DM interactions, which stabilizes antiskyrmions instead of skyrmions [88].

Indeed, the topological spin textures can also be stabilized by other mechanisms such as the frustrated exchange interactions [89], four-spin exchange interactions [8] as well as the long-range dipolar interactions [90-93]. However, most theoretical and experimental studies in the field focused on the systems with DM interactions during the last decade, as the DM interaction can be induced and adjusted by using modern interface engineering and multilayer fabrication techniques [2, 3, 5-7, 94-101]. As such, for the rest of this review, we will focus on the magnetic skyrmion textures stabilized by DM interactions, while other mechanisms will also be briefly reviewed in the last section. It is worth mentioning that, although conventional magnetic bubbles can be stabilized by the competition between ferromagnetic Heisenberg exchange and long-range dipolar interactions. The in-plane spin texture (i.e., helicity and chirality) of these common bubbles are not fixed, so that they cannot be treated as a topological object in the strict context of topological magnetism.

For the last decade, these topological spin textures have been significantly highlighted, mainly inspired by their potential to handle information in devices at low energy consumption and/or high processing speed. In particular, the study of magnetic skyrmions for the purpose of designing novel spintronic applications has led to an emerging research field called *skyrmion-electronics* [1-3], which is also referred to as *skyrmionics* in some contexts [5]. It is an important task to study and understand how to stabilize and manipulate these topological spin textures, to eventually realize spin texture-based spintronic applications, such as memories, logic computing elements and transistor-like functional devices. In the following, we will review the representative topological spin texture – the magnetic skyrmion, from the points of views of writing, deleting, reading and processing magnetic skyrmions toward spintronic applications.

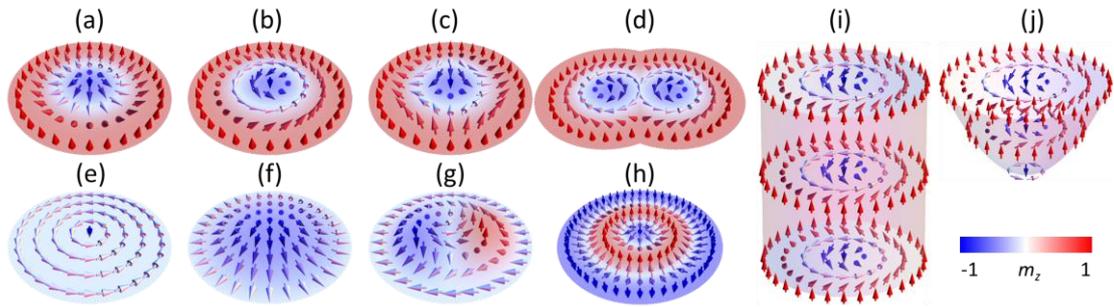

Figure 1. Illustrations of a series of 2D and 3D topological spin textures in magnetic materials. (a) Néel-type skyrmion ($Q = -1$), (b) Bloch-type skyrmion ($Q = -1$), (c) antiskyrmion ($Q = +1$), (d) biskyrmion ($Q = -2$), (e) vortex ($Q = -0.5$), (f) meron ($Q = -0.5$), (g) bimeron ($Q = -1$), (h) skyrmionium ($Q = 0$), (i) skyrmion tube, and (j) magnetic bobber. The arrow denotes the spin direction and the out-of-plane spin component ($m_z$) is represented by the color: red is out of the plane, white is in-plane, and blue is into the plane.

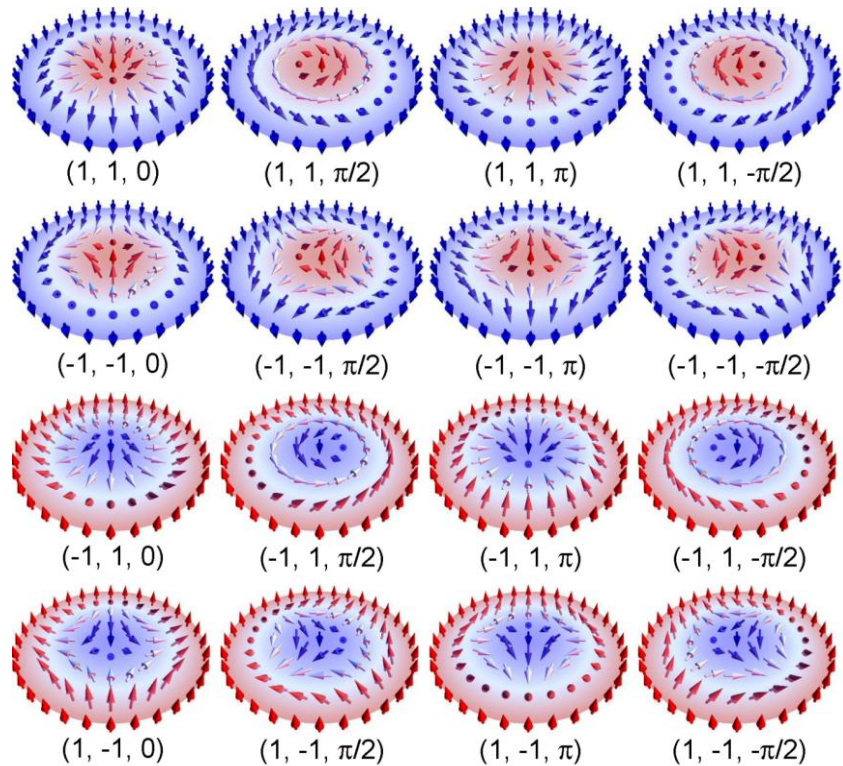

Figure 2. Illustrations of 2D magnetic skyrmions with different topological charge, vorticity number and helicity number, i.e. ($Q$, $Q_v$, $Q_h$). The arrow denotes the spin direction and the out-of-plane spin component ($m_z$) is represented by the color: red is out of the plane, white is in-plane, and blue is into the plane.

## 1.2. Magnetic Skyrmions at Low Temperature

In the history of magnetism, many new phenomena and magnetic structures were first discovered at low temperature due to the low transition temperature of magnetic materials, where the thermal fluctuations are largely reduced. Likely, ten years ago in 2009, Mühlbauer et al. for the first time experimentally observed the lattice structure of magnetic skyrmions in the chiral itinerant-electron magnet MnSi at a low temperature about 29 K [75] using a neutron scattering measurement [see Fig. 3(a)]. The lattice of magnetic skyrmions, so-called the *skyrmion lattice*, is a hexagonal ordered array of skyrmions and often referred to as the *skyrmion crystal*. As shown in Fig. 3(b), in 2010, such a hexagonal skyrmion crystal was directly observed in a thin film of crystalline $Fe_{0.5}Co_{0.5}Si$ at a low temperature of 25 K by Yu et al. [102], where the real-space imaging of magnetic spin textures was acquired by using the Lorentz transmission electron microscopy (LTEM). Using the same technique, in 2011, Yu et al. reported the formation of the skyrmion crystal in FeGe over a wide range of temperature ranging from 60 K to 260 K [103], which is very close to room temperature. In 2012, Yu et al. further demonstrated the current-induced motion of the skyrmion crystal in a FeGe thin film at a near room temperature, ranging from 250 K to 270 K [104]. In 2013, the transition of a skyrmion crystal phases to other magnetic textures (including helimagnetic, conical and ferromagnetic phases) in $Fe_{1-x}Co_xSi$ ($x = 0.5$) was also observed by using the magnetic force microscopy (MFM) at a very low temperature of 10 K [105]. Stabilization of skyrmions in such non-centrosymmetric crystalline materials is known to originate from the bulk-form DM interaction due to the broken inversion symmetry as discussed in Sec. 1.1.

However, as was discussed in earlier Sec. 1.1 of this article, the existence of magnetic skyrmions turns out to be possible not only in ferromagnetic bulk metals, but also ultrathin films where the interface-oriented DM interaction is harnessed. In 2011, Heinze et al. experimentally revealed the 2D square skyrmion crystal in a hexagonal monolayer crystalline Fe film grown on the Ir surface using the spin-polarized scanning tunnelling microscopy (SP-STM) at a low temperature of 11 K [8]. Such a topological spin texture stems from the interplay among the Heisenberg exchange, four-spin and DM interactions. In such interface-oriented DM interaction-governed system, Romming et al. for the first time experimentally realized both the writing and deleting of individual skyrmions in a PdFe bilayer on Ir(111) at a low temperature of 4.2 K [see Fig. 3(c)], where skyrmions are controlled by local spin-polarized currents from a STM. At the time, an out-of-plane magnetic field of several Tesla (e.g., B = ~3 T) is required to stabilize individual skyrmions at low temperatures [106]. It is worth to mention that Jonietz et al. [78] and Schulz et al [107] also reported the ultralow threshold current, $10^6$ A m$^{-2}$, for depinning magnetic skyrmions at ~26 K, which is 4-5 orders smaller than the depinning threshold of ferromagnetic domain walls, and these demonstrations excited great interest on using skyrmions for low power spintronic applications together with earlier findings. However, as skyrmions driven by the extremely low depinning current were displaced at a relatively low velocity [1, 107, 108], the required current density for actual devices may be increased as skyrmions need to move with a fast velocity.

At low temperatures, the hosting materials of skyrmions are not limited to conventional metallic ferromagnets, and other kinds of topological magnetic textures can also be stabilized. In 2012, Seki and colleagues [109] for the first time identified skyrmions in an insulating multiferroic magnet $Cu_2OSeO_3$ through LTEM and magnetic susceptibility measurements below the temperature of 60 K [see Fig. 3(d)]. By utilizing the small-angle neutron scattering (SANS), Seki et al. also found a triangular skyrmion lattice in multiferroic insulator $Cu_2OSeO_3$ [110].

In 2016, Matsumoto et al. directly observed skyrmion domain boundaries in $FeGe_{1-x}Si_x$ ($x$ ~ 0.25) using the differential phase contrast scanning transmission electron microscopy (DPC STEM) equipped with a segmented annular all-field detector at a temperature of 95 K [111], where they found that individual skyrmions at the domain boundary cores can flexibly change their size and shape to stabilize their core structures. The flexibility of the skyrmion structure at domain boundaries could be a basis for creating twisted skyrmion [112] that can be used for applications. Here it is worth mentioning that the DPC STEM technique can directly and precisely observe electromagnetic structures at an ultra-high resolution down to 1 nm. In 2016, McGrouther et al. revealed the internal spin structures of skyrmions in a freestanding nanowedge specimen of cubic B20 structured FeGe by using the DPC STEM technique but equipped with a pixelated detector [113]. In 2018, also by using the DPC STEM with a pixelated detector, McVitie et al. imaged the Néel-type domain wall structure and precisely measured the domain wall width [114].

These early stage experimental observations of skyrmions at low temperatures mainly focused on the existence of skyrmion crystals, skyrmion chain, skyrmion cluster, and individual isolated skyrmions in diverse material platforms stabilized by either bulk or interfacial DM interactions. Although these findings have proved fascinating physical properties of skyrmions and provided in-depth physical understanding behind their topological characteristics, from the viewpoint of practical applications, most commercial electronic devices based on skyrmions would require their room-temperature stabilization (or at even higher temperature). In the following, we will review the discovery of skyrmions and relevant topological spin textures at room temperature.

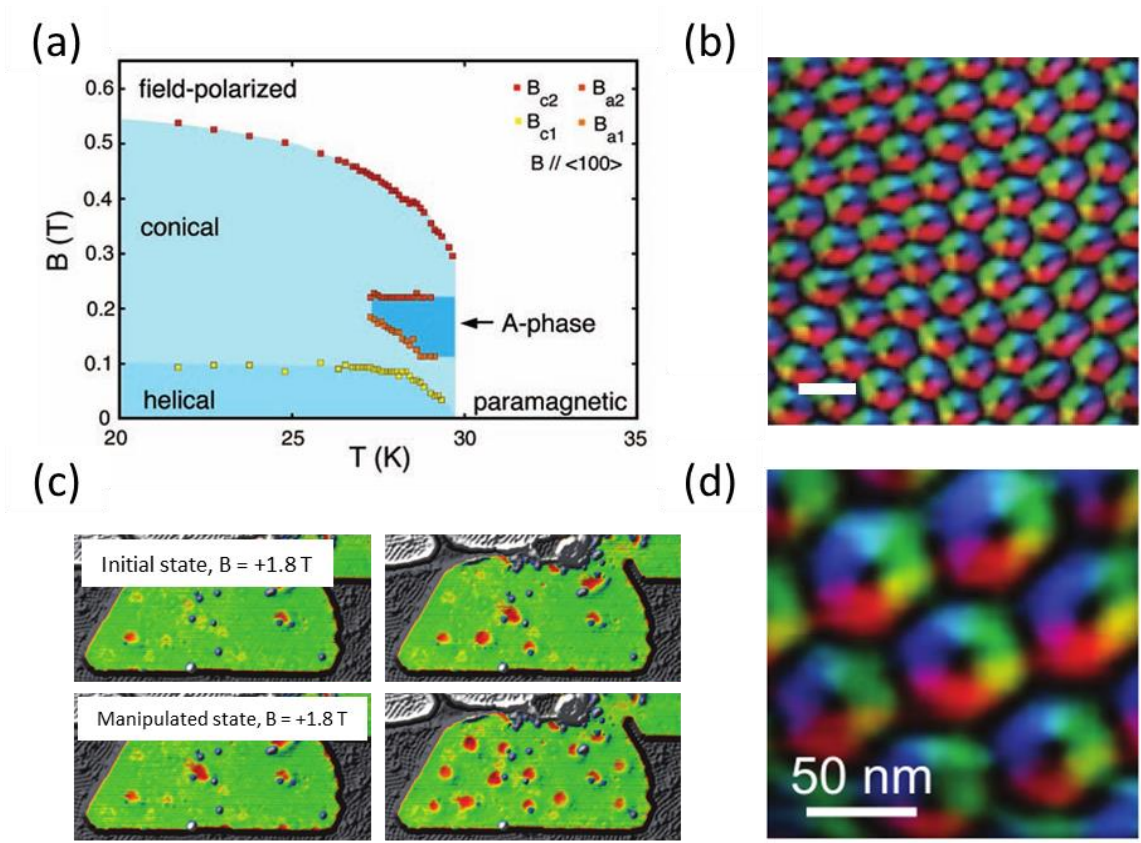

Figure 3. Observation of magnetic skyrmions at low temperature. (a) Magnetic phase diagram of MnSi. For $B = 0$, helimagnetic order develops below $T_c = 29.5$ K. Reprinted with permission from [75]. Copyright © 2009 AAAS. (b) A hexagonal skyrmion crystal observed using Lorentz TEM in a thin film of crystalline $Fe_{0.5}Co_{0.5}Si$ at a low temperature of 25 K. Scale bar, 100 nm. Reprinted with permission from [102]. Copyright © 2010 Macmillan Publishers Limited. (c) Successive population of the island with skyrmions by injecting higher-energy electrons through local voltage sweeps. Reprinted with permission from [106]. Copyright © 2013 AAAS. (d) Skyrmion crystal in $Cu_2OSeO_3$ thin film obtained through the analysis of Lorentz TEM data taken at 5K. Reprinted with permission from [109]. Copyright © 2012 AAAS.

## 1.3. Discovery of Room-Temperature Skyrmions

Between 2015 and 2016, there have been several reports of room-temperature stabilized magnetic skyrmions. The most notable demonstration was the stabilization of room-temperature skyrmions in the industry-relevant sputter-grown non-crystalline ferromagnetic heterostructures. This approach was motivated by the theoretical suggestion [9] and the experimental confirmation [13, 115] of homochiral Néel-type domain walls in such ultrathin ferromagnet-based asymmetric heterostructures at room temperature. In 2015, Jiang et al. not only observed stable room-temperature individual skyrmion bubbles in a Ta (5 nm)/CoFeB (1.1 nm)/TaO$_x$ (3 nm) trilayer grown by magnetron sputtering [see Fig. 4(a)], but also demonstrated the current-driven transformation of skyrmion bubbles from strip domains in such a trilayer device via a geometrical constriction [116]. It should be noted that the term '*skyrmion bubble*' studied in this report is a topologically nontrivial object with a skyrmion number of $|Q| = 1$ and a fixed Néel-type chirality induced by the DM interaction. However, unlike the compact skyrmion with 1-100 nm diameter, the skyrmion bubble has a larger size (~ 1 micrometer) comparable to trivial achiral bubbles stabilized by dipolar fields [91, 117-119]. Both skyrmion bubble and compact skyrmion are topological spin textures and turned out to be equivalent in the topological definition [see Eq. (3)], but they are often differentiated with two different names – skyrmion bubble or compact skyrmion – in some contexts due to their significant size difference. Soon after this report, in 2016, some other groups subsequently reported the observation of room-temperature skyrmions in similar ultrathin ferromagnetic asymmetric heterostructures.

Moreau-Luchaire et al. fabricated the Ir/Co/Pt asymmetric multilayers [29], where an additive interface-induced DM interaction of about 2 mJ m$^{-2}$ was achieved, and using the scanning X-ray transmission microscopy (SXTM), sub-100 nm individual skyrmions were imaged in the asymmetric multilayers at room temperature and low magnetic fields (e.g., $B_z = 58$ mT) [see Fig. 4(b)]. As the thermal stability of magnetization increases with the volume of the magnet, the skyrmions in the ten repetitions of the Ir/Co/Pt films are stable against thermal fluctuations at room temperature [29]. As shown in Fig. 4(c), Boulle et al. also observed stable skyrmions in sputtered ultrathin Pt/Co/MgO nanostructures at room temperature and zero external magnetic field [120]. Based on the in-plane magnetization sensitive X-ray magnetic circular dichroism photoemission electron microscopy (XMCD-PEEM) experiments, it was found that the skyrmions in the Pt/Co/MgO thin films have the left-handed chiral Néel structure [120], which indicates they are stabilized by interface-induced DM interactions.

In the same year, Woo et al. reported the observation of stable room-temperature skyrmions and their current-induced motion in [Pt (3 nm)/Co (0.9 nm)/Ta (4 nm)]$_{15}$ and [Pt (4.5 nm)/CoFeB (0.7 nm)/MgO (1.4 nm)]$_{15}$ multilayer stacks by using both the full-field magnetic transmission soft X-ray microscopy (MTXM) and STXM [121]. Woo et al. demonstrated the stabilization of skyrmion lattice in a confined circular magnetic disk of about 2-μm diameter [see Fig. 4(d)], and further reported the motion of a train of individual skyrmions at a speed up to 100 m s$^{-1}$ driven by a current density of $j \sim 5 \times 10^{11}$ A m$^{-2}$ [121]. This is a direct experimental evidence that suggests skyrmions can be used for real-world high-speed spintronic applications. Later in 2016, as shown in Fig. 4(e), Yu et al. also experimentally demonstrated the creation of room-temperature skyrmion bubbles in ultrathin CoFeB films [122], where skyrmion bubbles are stabilized by a delicate competition among the Heisenberg exchange interaction, dipolar interaction, and perpendicular magnetic anisotropy (PMA). The skyrmion bubbles found in this work have a skyrmion number of $|Q|$ = 1 and typical size close to 1 micrometer. Yu et al. reported that, although the DM interaction strength is only of 0.25 mJ m$^{-2}$ in the CoFeB films, it is strong enough to fix the Néel-type chirality of skyrmion bubbles, leading to the Néel-type chiral nature. In 2017, Soumyanarayanan et al. [31] found a tunable room-temperature skyrmion platform in the Ir/Fe/Co/Pt multilayer stacks for studying sub-50-nm skyrmions [see Fig. 4(f)], where one can adjust the magnetic interactions governing skyrmions by varying the ferromagnetic layer composition. In 2018, Zhang et al. experimentally realized the control of skyrmion density in Pt/Co/Ta multilayer stacks by varying temperature [123] or in-plane magnetic field [124].

The existence of room-temperature magnetic skyrmions were also demonstrated using non-centrosymmetric crystalline interfaces and bulk materials. As shown in Fig. 4(g), Chen et al. imaged the room-temperature magnetic skyrmions in the Fe/Ni/Cu/Ni/Cu(001) multilayers using the spin-polarized low-energy electron microscopy (SPLEEM) [125], where the exchange coupling across non-magnetic spacer layers is responsible for the stabilization of skyrmions. A new class of cubic chiral magnets hosting room-temperature skyrmion crystals – $\beta$-Mn-type Co-Zn-Mn alloys, was found by Tokunaga and colleagues [126]. The room-temperature magnetic skyrmion crystal was also realized in an artificial manner, as demonstrated by Gilbert et al. via patterning asymmetric magnetic nanodots with controlled circularity on an underlayer with PMA [127]. Such an approach is theoretically predicted by Sun et al. [128], which will be discussed in Sec. 2 as a possible method for writing skyrmions.

As mentioned in Sec. 1.1, magnetic skyrmions can be stabilized in different materials by different mechanisms. The room-temperature skyrmions in asymmetric ferromagnetic multilayer stacks are usually stabilized by enhanced interface-induced DM interactions. In 2017, Hou et al. [129] for the first time observed the skyrmion bubbles with variable topological charges at room temperature in the frustrated kagome $Fe_3Sn_2$ magnet with uniaxial magnetic anisotropy, by using in-situ LTEM. In the same year, antiskyrmions with $Q_v$ = -1 were experimentally identified in acentric tetragonal Heusler compounds $Mn_{1.4}Pt_{0.9}Pd_{0.1}Sn$ by Nayak et al. [88]. These two works not only provide more material platforms for hosting magnetic skyrmions and other topological spin textures, but also prove that topological spin textures with different topological charges can be stabilized at room temperature, shedding light on possible future applications based on different topological spin textures.

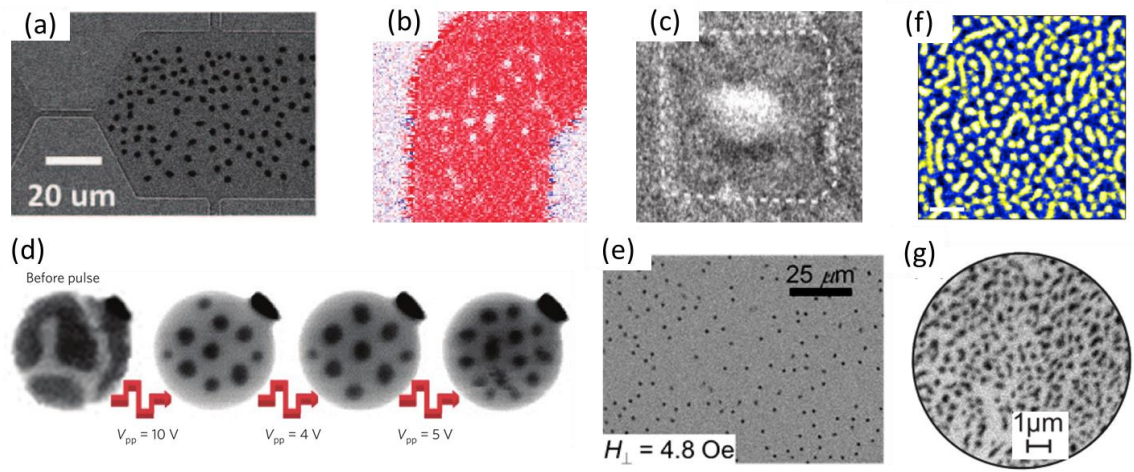

Figure 4. Observation of magnetic skyrmions at room temperature. (a) The current-induced generation of skyrmion bubbles in Ta (5 nm)/$Co_{20}Fe_{60}B_{20}$ (1.1 nm)/$TaO_x$ (3 nm). Reprinted with permission from [116]. Copyright © 2015 AAAS. (b) A 1.5×1.5 $\mu m^2$ out-of-plane magnetization ($m_z$) map obtained by STXM on a $(Ir/Co/Pt)_{10}$ multilayer with applying out-of-plane magnetic field of 68 mT. Reprinted with permission from [29]. Copyright © 2016 Macmillan Publishers Limited. (c) XMCD-PEEM image of a 420 nm square dot (indicated by the dotted line) which is patterned with Ta (3 nm)/Pt (3 nm)/Co (0.5–1 nm)/$MgO_x$/Ta (2 nm) thin film. Reprinted with permission from [120]. Copyright © 2016 Macmillan Publishers Limited. (d) An initial labyrinth domain state was generated by static field (first image) and then transformed into a hexagonal skyrmion lattice in a 2μm Pt/Co/Ta disc by applying a bipolar pulse train with $V_{pp}$ = 10V (second image). The last two images were acquired after applying $V_{pp}$ =4V and $V_{pp}$ = 5V, respectively. Dark (light) contrast corresponds to up (down) magnetization in all STXM images except for the last three, where the X-ray magnetic circular dichroism (XMCD) contrast was inverted. Reprinted with permission from [121]. Copyright © 2016 Macmillan Publishers Limited. (e) Polar-MOKE images of CoFeB thin film with PMA field of $H_k$ ≈ 1.1 kOe for 4.8 Oe out-of-plane external magnetic field.

Reprinted with permission from [122]. Copyright © 2016 American Chemical Society. (f) Microscopic imaging (scale bar, 0.5 μm) of sample Fe (3)/Co (6) (number of atomic layers in braces) with MFM at ~ −100 mT of applied field. Reprinted with permission from [31]. Copyright © 2016 Macmillan Publishers Limited. (g) The out-of-plane SPLEEM images of Fe (2.6 ML)/Ni (2 ML)/Cu (8.6 ML)/Ni (15 ML)/Cu (001) multilayer structures (1 ML ~ 1.8 Å). Reprinted with permission from [125]. Copyright © 2015 AIP Publishing.

## 1.4. Potential Applications

The field of *skyrmion-electronics* mainly focuses on the design and development of information storage and processing devices [2, 3, 5-7, 130]. The most highlighted skyrmionic information storage device – skyrmionic racetrack memory is proposed by Fert et al. in 2013 [1], which is an improved design based on the domain-wall racetrack memory proposed by Parkin et al. in 2008 [131]. Later in 2013, Iwasaki et al. [108, 132] and Sampaio et al. [133] independently studied the current-induced dynamics of isolated skyrmions in ferromagnetic racetracks, which provided fundamental insight into the design of skyrmion-based racetrack memory. Since 2013, a number of theoretical and experimental works were carried out on the skyrmion-based racetrack memory and its derived applications, however, the experimental demonstration of such fully functional electrically operating skyrmion-based memory device is awaiting its demonstration.

In 2013, the magnetoelectric resonance effect of magnetic skyrmions was found by Mochizuki et al. [134] and Okamura et al. [135], which suggests an opportunity for building the skyrmion-based microwave applications, such as microwave detector [136]. Besides, the skyrmion-based nano-oscillators [137-139] can be built using the oscillation characteristics induced by currents.

In 2015, Zhang et al. [140] theoretically proposed the prototype of skyrmion-based logic computing gates, which can perform the logic AND and OR operations. Zhang et al. [141] also proposed a transistor-like functional device based on skyrmions, where the current-induced motion of skyrmions is controlled by a gate voltage. Note that a similar idea on the control of skyrmion dynamics by a gate voltage was also proposed by Upadhyaya et al. [142] in the same year. These potential skyrmion-based information computing devices can be combined with the skyrmion-based racetrack memory devices, which may lead to the invention of an in-memory logic computing circuit based on skyrmions.

Some bio-inspired applications based on skyrmions were also proposed in recent two years. Huang et al. [143] designed a skyrmion-based artificial synapse device for neuromorphic computing systems in 2017, which was recently experimentally realized by Song et al. [144]. Li et al. [145] also proposed a skyrmion-based artificial neuron that mimics the leaky-integrate-fire function of a biological neuron. In 2018, Bourianoff et al. [146] and Prychynenko et al. [147] proposed the first reservoir computing model based on magnetic skyrmions. Most recently, the thermally activated dynamics of skyrmions was investigated by several groups and found to be useful for computing [148-159]. For example, the thermal skyrmion diffusion-based signal reshuffling device operation was experimentally realized by Zázvorka et al. [160], which could be used in future skyrmion-based probabilistic computing devices. Besides, Nozaki et al. experimentally realized the control of the thermal diffusion of skyrmions by applying a pure voltage [149].

For the successful demonstration of any skyrmion-based potential applications, the energy-efficient writing, deleting, reading and processing processes of skyrmions are prerequisites. In the following, we review the progresses that have been made over the last 10 years in the writing, deleting, and reading of magnetic skyrmions. We will also review several representative skyrmion-based applications in detail, including the skyrmion-based racetrack memory, logic computing gate, transistor-like device and bio-inspired applications.

## 2. Writing and Deleting Skyrmions

Magnetic skyrmion is a promising block for building future spintronic applications, as it can be employed as a non-volatile information carrier in magnetic media. The simplest method to encode binary information using magnetic skyrmions in magnetic materials is based on the writing and deleting of individual isolated skyrmions. Namely, the presence of a skyrmion can stand for the binary information digit "1" or "0". Hence, the controllable and reliable creation and annihilation of skyrmions are prerequisites for any skyrmion-based information storage applications. It is a vital task to find the best method to write and delete skyrmions.

### 2.1. Magnetic Field

Magnetic field is one of the most important and controllable external stimuli that can be rather easily realized in most laboratories. In the early stage of skyrmion research, a number of theoretical and experimental studies were carried out on the magnetic phase diagram, where skyrmion crystals and isolated skyrmions usually form in chiral magnets within the range of certain out-of-plane magnetic fields and temperature [54, 75, 78, 102, 103, 105, 109, 161-163]. That is to say, by applying an external magnetic field perpendicularly to the magnetic film with DM interactions, it is possible to control the formation of skyrmion textures [106]. Some of the initial demonstrations of room-temperature skyrmions introduced in Sec. 1.1 were also realized by shrinking maze stripe domains into circular bubble-shaped domains by applying magnetic field.

On the other hand, once the skyrmion is created in a magnetic film, its size also depends on the strength of the external out-of-plane magnetic field [11, 164, 165] that interplays with other energetic contributions e.g. DM interaction, magnetic anisotropy, demagnetization energy and exchange interaction. Namely, a small out-of-plane magnetic field is usually required for the formation of skyrmions, while a large out-of-plane magnetic field could lead to the collapse and annihilation of skyrmions. In 2016, Müller et al. predicted and numerically demonstrated that, by applying magnetic field pulses, skyrmions can be created close to the edge of a chiral magnet due to the edge instability [166]. In 2017, M. Mochizuki [167] demonstrated that in a magnetic film with a fabricated hole or notch, individual skyrmions can be created by an external magnetic field in a controlled manner [see Fig. 5(a)]. Most recently, the writing of skyrmions in a uniformly magnetized film by a magnetic dipole was theoretically investigated by Garanin et al. [168], which can be seen as the creation of skyrmions by a localized magnetic field. In 2018, such a creation method of writing skyrmions was experimentally demonstrated by Zhang et al. using a scanning local magnetic field provided by magnetic force microscopy (MFM) tips [169].

Once the magnetic skyrmion is created, it can also be manipulated or even deleted by magnetic fields, similar to the magnetic vortices in magnetic disks [170, 171]. For example, Büttner et al. showed that the gigahertz gyrotropic eigenmode dynamics of a single skyrmion bubble can be excited by an external magnetic field pulse and corresponding magnetic field gradient [172]. Magnetic skyrmions in nanostructures with confined geometries can also be switched by microwave magnetic fields [see Fig. 5(b)] [173] or magnetic field pulses [see Fig. 5(c)] [174, 175]. Note that the microwave magnetic field can also excited other dynamics of skyrmions [176, 177]. Indeed, when the strength of magnetic field is larger than certain threshold, it is destructive for skyrmions and may ultimately lead to the fully polarized magnetic state, i.e. the ferromagnetic state. Hence, the application of a strong magnetic field, either in-plane or out-of-plane, is also the easiest way to delete skyrmions from magnetic media.

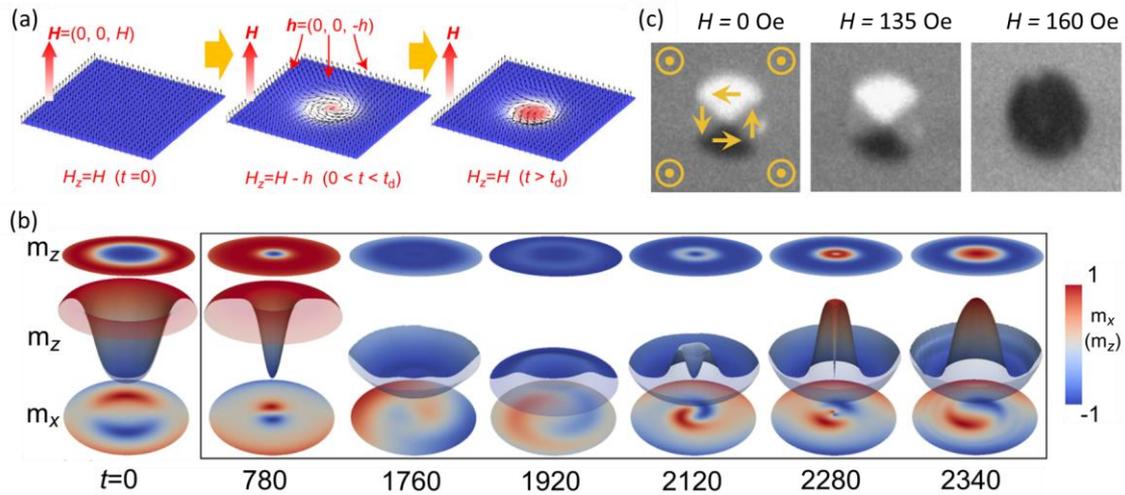

Figure 5. Writing, deleting and switching of skyrmions using magnetic fields. (a) Schematic of skyrmion creation in a magnetic field application. Reprinted from [167]. With permission of AIP Publishing. (b) The skyrmion core reversal induced by an AC magnetic field pulse. Reprinted with permission from [173]. Copyright © 2015 AIP Publishing. (c) The skyrmion core annihilation process with applying an in-plane magnetic field pulse. The yellow symbols indicate the out-of-plane spins. Reprinted with permission from [174]. Copyright © 2014 Macmillan Publishers Limited.

## 2.2. Spin-Polarized Electric Current

Spin-polarized electric current could provide spin-transfer torques (STTs) or spin-orbit torques (SOTs) on magnetic moments, which is an important driving force for magnetization dynamics in modern spintronic devices [97, 178, 179]. In 2011, the interaction between STTs and skyrmion lattices in chiral magnets was theoretically studied by Everschor et al. [180]. Then, the writing of a magnetic skyrmion using the spin-polarized electric current was theoretically predicted by Tchoe et al. in 2012, where skyrmions can be nucleated from a ferromagnetic background by a vertical current injection with the current density of the estimated order of $j \sim 10^{10}$-$10^{11}$ A m$^{-2}$ [181]. In 2013, Iwasaki et al. [132] and Sampaio et al. [133] independently studied the current-induced skyrmion creation in confined geometries using theoretical and numerical approaches. In particular, Iwasaki et al. numerically demonstrated that skyrmions can be nucleated and created from the notch in a nanotrack [132], which is an effective method to write skyrmions at a certain location. In the same year, as was noted in Sec. 2.1, Romming et al. [106] for the first time experimentally realized both the writing and deleting of single isolated magnetic skyrmions by using the local tunneling spin-polarized currents (~ 1 nA) from a STM at low temperature [see Fig. 6(a)].

In 2014, Zhou and Ezawa theoretically predicted that a magnetic skyrmion can be converted from a domain-wall pair in a junction geometry ($j \sim 1.8 \times 10^{12}$ A m$^{-2}$) [182]. In a similar manner, in 2015, Jiang et al. demonstrated in a room-temperature MOKE experiments that magnetic skyrmion bubbles can be created by utilizing the spin current-driven ($j \sim 0.5 \times 10^{10}$ A m$^{-2}$) transformation between strip domains and skyrmion bubbles in a Ta/Co$_{20}$Fe$_{60}$B$_{20}$/TaO$_x$ trilayer with a junction constriction [116]. The creation of skyrmion bubbles in such a junction constriction driven by inhomogeneous spin currents was further theoretically studied independently by Heinonen et al. [183], S.-Z. Lin [184], and Liu et al. [185], confirming its efficient mechanism and device compatibility suitable for future applications.

The spin-polarized current can also result in the creation of skyrmions or skyrmion bubbles by other different mechanisms. In 2016, Yuan and Wang theoretically demonstrated that a skyrmion can be created in a ferromagnetic nanodisk by applying a nano-second current pulse ($j \sim 2.0 \times 10^{12}$ A m$^{-2}$) [186]. Yin et al. also suggested in a theoretical study that it is possible to create a single skyrmion in helimagnetic thin films using the dynamical excitations induced by the Oersted field and the STT given by a vertically injected spin-polarized current ($j \sim 1.7 \times 10^{11}$ A m$^{-2}$) [187]. In 2017, as shown in Fig. 6(b), Legrand et al. experimentally realized the creation of magnetic skyrmions by applying a uniform spin current directly into nanotracks ($j \sim 2.38 \times 10^{11}$ A m$^{-2}$) [32]. Latter in 2017, Woo et al. experimentally demonstrated the creation of skyrmions at room temperature and zero external magnetic field by applying bipolar spin current pulses ($j \sim 1.6 \times 10^{11}$ A m$^{-2}$) directly into a Pt/CoFeB/MgO multilayer [33], which turned out to be thermally-induced skyrmion generation as systematically studied recently by Lemesh et al. [34]. Hrabec et al. also reported the experimental creation of skyrmions induced by applying electric current ($j \sim 2.8 \times 10^{11}$ A m$^{-2}$) through an electric contact placed upon a symmetric magnetic bilayer system [188]. The spin-polarized current-induced generation of skyrmions can be deterministic and systematic when pinning sites or patterned notches, where PMA is locally reduced, are used as the source of skyrmion generation as reported by Buttner et al. [35] in 2017 ($j \sim 2.6 \times 10^{11}$ A m$^{-2}$) [see Fig. 6(c) and Woo et al. [36] in 2018 ($j \sim 2.5 \times 10^{10}$ A m$^{-2}$) [Fig. 6(d)]. More recently, Finizio et al. reported that localized strong thermal fluctuation could also introduce systematic skyrmion generation at a designed location [189].

On the other hand, the spin-polarized current can delete skyrmions in nanostructures. For example, in 2017, De Lucia et al. theoretically studied the annihilation of skyrmions induced by spin current pulses ($j \sim 5 \times 10^{12}$ A m$^{-2}$) and suggested that skyrmions can be reliably deleted by designing the pulse shape [190]. In 2018, Woo et al experimentally demonstrated such deterministic deletion of a single skyrmion in ferrimagnetic GdFeCo films with the application of designed current pulses ($j \sim 2.5 \times 10^{10}$ A m$^{-2}$) [36]. In Table 1, we summarized the typical current density applied to create and delete skyrmions in the above mentioned theoretical and experimental works for the purposes of reference and comparison.

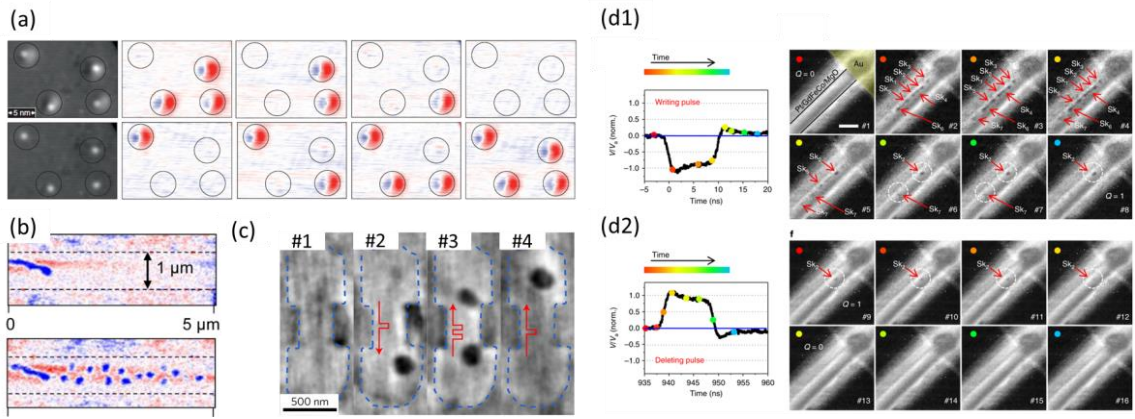

Figure 6. Writing and deleting of skyrmions using spin-polarized electric currents. (a) Creation and annihilation of single skyrmions with local spin-polarized currents at T = 4.2 K in the PdFe bilayer. Reprinted with permission from [106]. (b) MFM images of the 1 μm wide tracks before any pulses and after the injection of 1000 current pulses of 200 ns in 0.6 nm thick Co multilayers. Reprinted with permission from [32]. (c) Single-skyrmion generation and subsequent motion in Pt/CoFeB/MgO multilayers. Reprinted with permission from [35]. Copyright © 2017 Macmillan Publishers Limited. (d) Magnetic skyrmion configuration for writing (d1) and deleting (d2). Scale bar, 500 nm. Reprinted with permission from [36].

**Table 1. Writing and deleting of skyrmions using spin-polarized electric currents with different current densities in both theoretical and experimental works.**

| Materials | Current density | Pulse width | $T$ (K) | B (mT) | Reference |
|---|---|---|---|---|---|
| Experimental Writing | | | | | |
| PdFe bilayer | 1~200 nA | | 4.2 | 1800 | [106] |
| Ta/Co$_{20}$Fe$_{60}$B$_{20}$/TaO | $0.5 \times 10^{10}$ A/m$^2$ | 1 $\mu$s (single) | RT | 0.5 | [116] |
| Ta/Co/[Pt/Ir/Co/]$_{10}$/Pt | $2.38 \times 10^{11}$ A/m$^2$ | 200 ns (1000 pulses) | RT | 20.5 | [32] |
| [Pt/Co$_4$Fe$_4$B$_2$/MgO]$_{20}$ | $1.6 \times 10^{11}$ A/m$^2$ | 20 ns (3.33 MHz, 5s) | RT | 0 | [33] |
| [Pt/CoFeB/MgO]$_{15}$ | $3.0 \times 10^{11}$ A/m$^2$ | 7.5 ns (10000 pulses) | RT | 40 | [34] |
| Pt/Ni/Co/Ni/Au/Ni/Co/Ni/Pt | $2.8 \times 10^{11}$ A/m$^2$ | 7 ns (single) | RT | 6 | [188] |
| [Pt/Co$_{60}$Fe$_{20}$B$_{20}$/MgO]$_{15}$ | $2.6 \times 10^{11}$ A/m$^2$ | 12 ns (single) | RT | 6.1 | [35] |
| [Pt/Gd$_{25}$Fe$_{65.6}$Co$_{9.4}$/MgO]$_{20}$ | $2.5 \times 10^{10}$ A/m$^2$ | 10 ns (single) | RT | 130 | [36] |
| Experimental Deleting | | | | | |
| [Pt/Gd$_{25}$Fe$_{65.6}$Co$_{9.4}$/MgO]$_{20}$ | $2.5 \times 10^{10}$ A/m$^2$ | 10 ns (single) | RT | 130 | [36] |
| Theoretical Writing | | | | | |
| CoPt | $1.8 \times 10^{12}$ A/m$^2$ | | 0 | 0 | [182] |
| CoFeB | $1.0 \times 10^{12}$ A/m$^2$ | 40 ns (single) | 0 | 0.5 | [183] |
| Co$_{20}$Fe$_{60}$B$_{20}$ | $1.5$~$3.0 \times 10^{10}$ A/m$^2$ | | 0 | 70 | [185] |
| CoPt | $2.0 \times 10^{12}$ A/m$^2$ | 0.7 ns (single) | 0 | 0 | [186] |
| | $1.7 \times 10^{11}$ A/m$^2$ | | 50 | | [187] |
| | $3.6 \times 10^{11}$ A/m$^2$ | | 0 | | [132] |
| Theoretical Deleting | | | | | |
| CoFeB | $5$~$20 \times 10^{12}$ A/m$^2$ | | 0 | 0 | [190] |

RT, room temperature.

## 2.3. Local Electric Field

In the last decade, a number of studies have been performed to explore skyrmion dynamics induced by electric current, however, the electric current often accompanies Joule heating that leads to energy losses or even permanent damages to metallic devices. Therefore, it may be necessary to find more reliable ways to drive skyrmion dynamics, and in this context, the pure electric field is one of the desired ways to drive skyrmion dynamics because of its ultralow power consumption with negligible Joule heating. A number of reports [109, 134, 135, 191, 192] had shown the possibility to control skyrmions in magnetoelectric materials, such as the chiral magnetic insulator $Cu_2OSeO_3$, using a pure electric field (due to the magnetoelectric coupling). In 2015, Mochizuki and Watanabe [193] theoretically proposed the writing of isolated skyrmions in a multiferroic thin film by applying a local electric field [see Fig. 7(a)], which is realized through the magnetoelectric coupling effect in multiferroic compounds. In 2016, the electric-field-controlled transition between the skyrmion lattice phase and conical phase was experimentally realized in magnetoelectric chiral magnet $Cu_2OSeO_3$ [194].

In ferromagnetic materials, the writing and deleting of magnetic skyrmions can also be induced by a pure electric field, usually based on the effect of voltage control of magnetic anisotropy (VCMA) [142]. In 2017, as shown in Fig. 7(b), Hsu et al. experimentally demonstrated the reversible transition between the ferromagnetic texture and the magnetic skyrmion by applying a local electric field to the Fe triple layer on Ir(111) [195]. Such a local electric field, which can induce the writing and deleting of magnetic skyrmions, were realized by using the SP-STM. In the same year, the writing and deleting of skyrmion bubbles controlled by electric field in a Pt/Co/oxide trilayer was realized in an efficient and reproducible manner by Schott et al. at room temperature [196].

In 2018, Srivastava et al. experimentally demonstrated the electric field turning of DM interactions in a Ta/FeCoB/TaOx trilayer [197], which can be used as a method to further control the skyrmion chirality. Huang et al. also experimentally realized the writing of skyrmions in the magnetoelectric compound $Cu_2OSeO_3$ [192].

In 2019, Ma et al. designed and fabricated a Pt/CoNi/Pt/CoNi/Pt multilayer, which is sandwiched between the indium tin oxide (ITO)/$SiO_2$ bilayer and the glass substrate, in the form of racetracks where the thickness of the films had a slope [198]. In such a nanostructure, Ma et al. demonstrated that many skyrmion bubbles can be created (i.e., corresponding to the skyrmion writing) and directionally displaced about 10 micrometres by applying a pure electric field [see Fig. 7(c)]. When the electric field is removed, the skyrmion bubbles will be annihilated, which can be regarded as the deleting of skyrmion bubbles.

Similar to the manipulations of skyrmions by magnetic field and electric current, the polarity of skyrmions can also be switched by a pure electric field, in principle. For example, Bhattacharya et al. numerically demonstrated the electric-field-induced switching of a magnetic skyrmion in a magnetic tunnel junction (MTJ) structure [199]. On the other hand, the dynamics of skyrmions can also be controlled by a pure electric field, as experimentally demonstrated by Nozaki et al. in a W/FeB/Ir/MgO multilayer structure recently [149], the thermal motion of skyrmion bubbles can be controlled by a pure voltage.

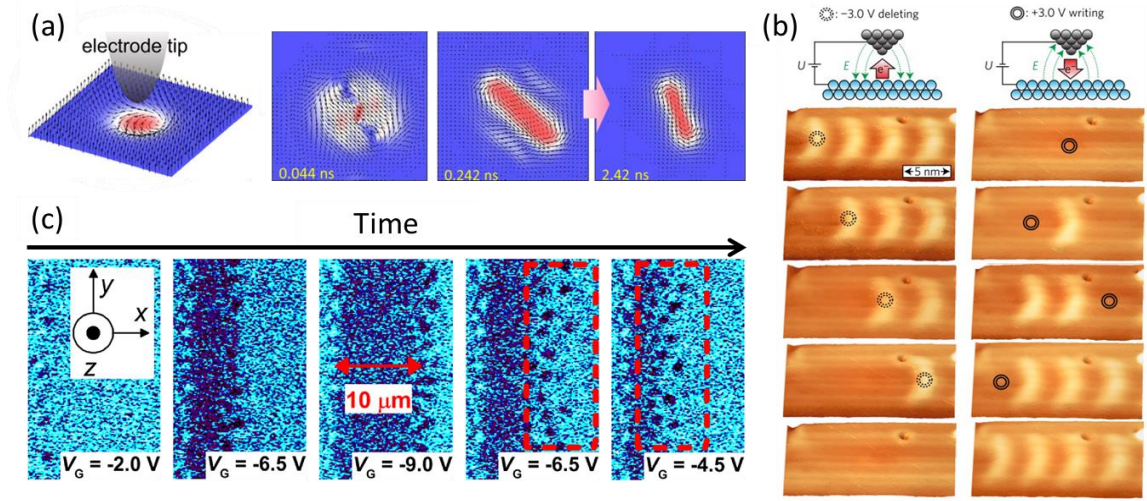

Figure 7. Writing and deleting of skyrmions by electric fields. (a) Schematic of skyrmion creation via local application of an electric field using an electrode tip and simulated spatiotemporal dynamics of the magnetizations for the electrical skyrmion creation process in the Cu$_2$OSeO$_3$ thin film. Reprinted with permission from [193]. (b) Perspective views of subsequent SP–STM constant-current images of the same Fe triple layer area showing the writing and deleting of individual magnetic skyrmions. Reprinted with permission from [195]. Copyright © 2017 Macmillan Publishers Limited. (c) MOKE microscopy images of the EF-induced creation and motion of a chiral domain wall accompanied by the creation of skyrmion bubbles in Pt/CoNi/Pt/CoNi/Pt multilayer with the thickness gradient at $H_z = -0.2$ mT. Reprinted with permission from [198].

## 2.4. Laser

The magnetization dynamics can also be driven by (ultrafast) laser pulses, and laser pulse-induced magnetization excitation has been of great interest for many years. Such excitation enabled the observation of ultrafast demagnetization [200] and all-optical switching [201-204], which could enable the integration of optical spintronics devices. Single laser shot-induced *deterministic* magnetization switching was also demonstrated using ferrimagnetic thin films, where the ultrafast last-induced heating could lead to the angular momentum transfer between two sublattices [202]. Unlike ferrimagnetic materials, the injected laser-induced local heating in ferromagnetic materials *probabilistically* lead to the switching of magnetization within the heated area due to the thermal perturbation and its interactions with e.g. local dipolar field [130, 205, 206]. In detail, the switching field (i.e. coercive field) of ferromagnets generally decreases with increasing temperature, therefore, the dipolar interactions that favor the formation of multi-domains overcome the switching field threshold, leading to the local magnetization switching. Nevertheless, in chiral magnetic materials, such laser-induced magnetization switching results in the local generation of a single or multiple chiral magnetic textures, including skyrmions. In 2009, Ogasawara et al. experimentally studied the magnetization reversal dynamics in a TbFeCo film induced by a femtosecond laser pulse in the presence of a small external magnetic field [207], where the laser-induced reversed magnetic domain has a size of about 1 micrometer. Such a circular magnetic domain, which is stabilized by the dipolar interaction, can be seen as a achiral skyrmion bubble [208]. In 2013, Finazzi et al. reported the creation of topological spin textures induced by ultrashort single laser pulses in an amorphous thin alloy film of $Tb_{22}Fe_{69}Co_9$ without the help of an external magnetic field [209], where both skyrmion and skyrmionium textures are created [see Fig. 8(a)]. The lateral size of the laser-created magnetic skyrmions can be as small as 150 nm [209].

In 2017, Fujita and Sato theoretically studied the ultrafast creation of skyrmion and skyrmionium textures by applying a vortex laser beam [68], where the switching of magnetization in either a ferromagnetic or antiferromagnetic background is a result of the laser-induced non-uniform temperature. In 2018, Yang et al. numerically studied the manipulation of skyrmions by an all-photonic orbital angular momentum transfer mechanism [210]. In the same year, as shown in Fig. 8(b), the laser-induced writing and deleting of skyrmions in the prototypical itinerant chiral magnet FeGe was experimentally realized by Berruto et al. [211], where the writing and deleting speed was found to be controlled by the cooling rate following the laser-induced temperature increase. Je et al. also experimentally demonstrated the laser-induced creation of disordered hexagonal skyrmion bubble lattices from a ferromagnetic background in an ultrathin $Ta/Fe_{72}Co_8B_{20}/TaO_x$ trilayer film at room temperature [see Fig. 8(c)], where the density of skyrmion bubbles was controlled by the laser fluence [206]. Such an experiment suggests that the skyrmion bubbles and bubble lattices can possibly be manipulated by lasers in a controlled manner.

Apart from the writing and deleting of skyrmions, magnetic skyrmions can be excited to different modes by optical methods. For example, Ogawa et al. reported the laser-induced collective excitation modes of the skyrmion phase in an insulating chiral magnet $Cu_2OSeO_3$ [212], including the rotation mode, breathing mode, and several additional spin precession modes.

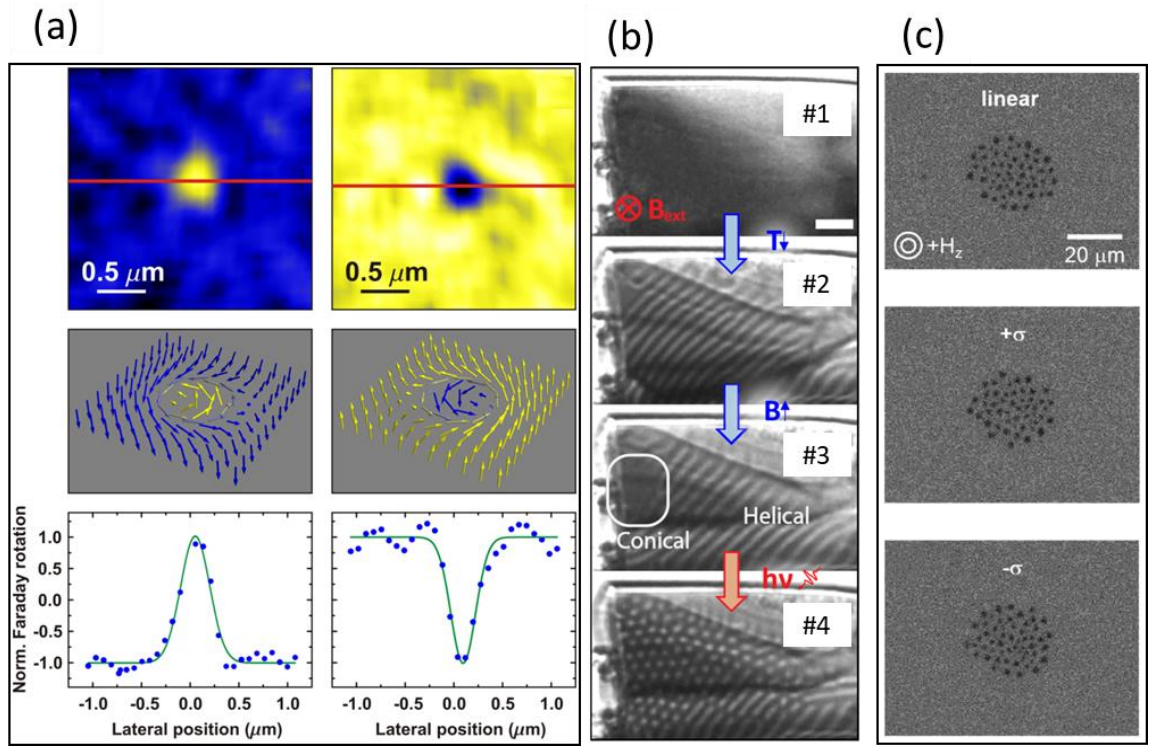

Figure 8. Writing of skyrmions by lasers. (a) Near-field Faraday rotation map showing magnetic domains induced in a thin TbFeCo film after single laser pulse irradiation, corresponding skyrmion spin textures, and profile (dots) of the Faraday rotation measured along the red lines. Reprinted with permission from [209]. (b) Lorentz-Fresnel micrographs of the FeGe nanoslab. The skyrmion lattice in #4 is created by near-IR fs laser pulses. Scale bar is 250 nm. Reprinted with permission from [211]. (c) Nucleated skyrmion bubbles from a saturated state ($+z$) by a 35-fs single laser pulse with a laser with linear polarization, right-handed $\sigma^+$ polarization, and left-handed $\sigma^-$ polarization. Reprinted with permission from [206].

## 2.5. Imprinting

Skyrmion-like magnetic vortex domains, which is stabilized by geometrical constrains, can also be created by means of nanoimprinting even in the absence of DM interactions. As shown in Fig. 9(a), in 2013, Sun et al. proposed a method for the creation of a skyrmion lattice based on a combination of a perpendicularly magnetized CoPt film and nanopatterned arrays of magnetic vortices that are geometrically confined within Co nanodisks [128]. Such a method is similar to the exchange spring model [213], where the spin textures in a soft phase (i.e., corresponding to the magnetic vortex in Co nanodisk) can be imprinted into the hard phase (i.e., corresponding to the CoPt film) due to the interlayer exchange coupling between the soft and hard phases. In principle, the created skyrmion lattice [see Fig. 9(b)] can be stabilized in a wide temperature and field range, even at room temperature, zero magnetic field, and in the absence of the DM interaction.

In 2015, Fraerman et al. experimentally observed the creation of skyrmions in a perpendicularly magnetized Co/Pt multilayer film via the imprinting method [214], where the Co/Pt multilayer film is exchange-coupled with a Co nanodisk with the vortex state. In the same year, by using the transmission soft X-ray microscopy, Streubel et al. experimentally studied the dynamics of imprinted skyrmion textures and found that skyrmionium textures can also be created by imprinting a magnetic vortex from the soft Py layer into the hard Co/Pd layer in the presence of reasonable interlayer exchange coupling [215]. The imprinting skyrmion lattice with controlled circularity and polarity as ground state at room temperature was also experimentally realized by Gilbert et al. [127], where the magnetic vortex state in Co nanodots were imprinted into the Co/Pd underlayer with PMA [see Fig. 9(c)]. On the other hand, Del-Valle et al. theoretically demonstrated the possibility to imprint skyrmions in thin films by superconducting vortices [216].

In 2018, Sun et al. numerically studied the stability and skyrmion Hall effect (SkHE) of magnetic skyrmions in the exchange-coupled system [217], which was used for the creation of skyrmions via imprinting effect in previous reports. They found that the stability of the skyrmion state can be enhanced due to the coupling between skyrmions in the thin film and the vortices in the capping nanodisks. At the same time, the capping nanodisks also act as attracting centers so that the SkHE can be effectively suppressed. The SkHE is a phenomenon related to the topology of skyrmion structures [26, 218, 219], which may be detrimental for the in-line motion of skyrmions. We will discuss the SkHE in detail in Sec. 4.

In this section, we have reviewed a series of methods for writing and deleting skyrmions in magnetic materials. From the viewpoint of practical applications, it is desirable to use the spin-polarized electric current to control the generation/deletion of skyrmions, which can be integrated into transistor-based conventional complimentary metal-oxide semiconductor (CMOS) scheme. Spin currents can be rather easily generated either by using a polarizer layer via spin-filtering effect or by using high spin-orbit coupling material with relatively large spin Hall effect (or interfaces with large Rashba effect), which have been extensively explored for the last few years. Such a heavy metal/ferromagnet heterostructure are also known to generate certain DM interaction and PMA to stabilize skyrmions. However, the method utilizing electric excitation often suffers from inevitable current-induced Joule heating and leads to a large power consumption, therefore the electric field and laser could also be alternative methods for writing and deleting skyrmions in a low-energy manner in the future.

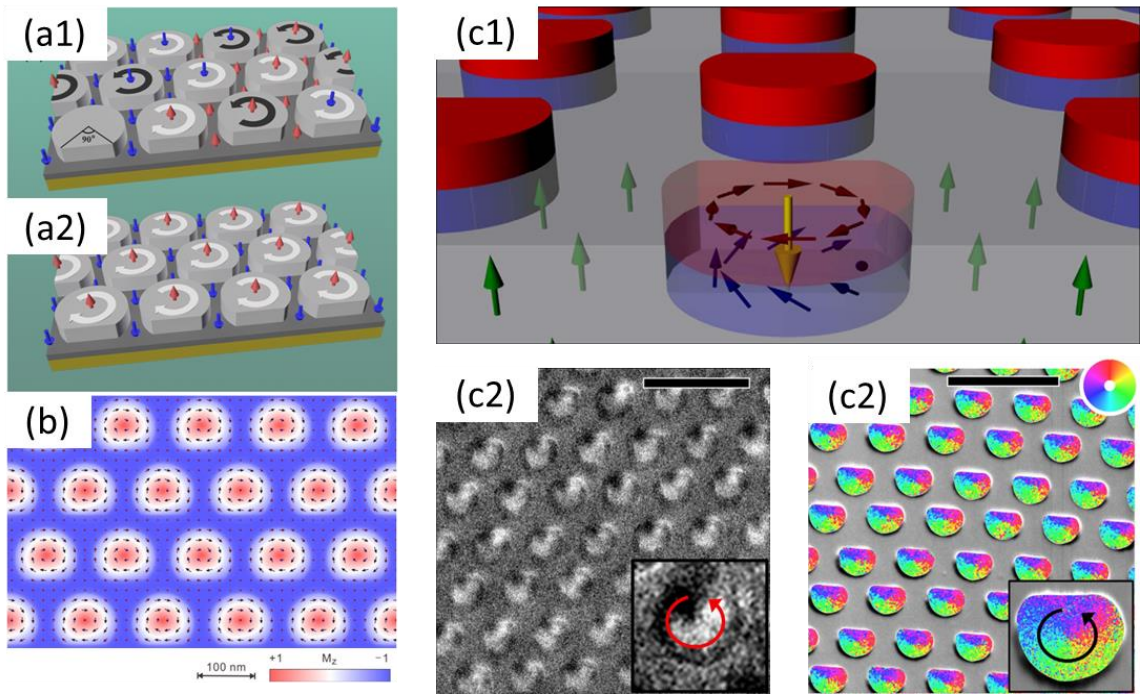

Figure 9. Writing of skyrmions by imprinting. (a)-(b) The theoretical proposal for creating the 2D skyrmion crystal. (a1) Ordered arrays of magnetic submicron disks are prepared on top of a film with perpendicular anisotropy. The arrows represent the magnetization orientation of the local moments. (a2) Skyrmion lattice creation with the field treatment. (b) Top view of magnetic configuration of the CoPt layer in a calculated artificial skyrmion crystal. (a)-(b) are reprinted with permission from [128]. (c) Imprinted artificial skyrmion lattices at room temperature. (c1) Illustration of the hybrid structure consists of Co dots (red) on top of Co/Pd PMA underlayer (grey) where the in-plane spin texture of the Co dots (purple arrows) is imprinted into an irradiated Co/Pd region (light blue) underneath the dots (tilted blue arrows). Remanent-state (c1) MFM and (c2) SEMPA (superimposed onto a scanning electron microscopy image of the dots) images, after saturating the dots in an in-plane field parallel to the flat edge of the dots to the right, indicate circularity control. Scale bar, 2 mm. A key to the magnetization winding direction is shown in the insets. Reprinted with permission from [127].

## 3. Reading Skyrmions

## 3.1. Microscopy Imaging of Magnetic Skyrmions: X-ray, TEM, SPM, MOKE

The manipulation of magnetic skyrmions as information carriers in device applications requires the precise tracking of their locations and dynamic behaviors. Hence, the efficient reading of skyrmions in real space is an important task that needs to be realized in a reliable manner with high enough on/off ratio. A lot of state-of-art microscopy magnetic imaging techniques have been used to observe the real-space profiles of skyrmions in laboratories for the purpose of studying their static and dynamic properties (see Table 2) [29, 33, 35, 36, 102, 106, 109, 111, 116, 120, 121, 125, 148, 160, 174, 195, 197, 220-233]. However, as a part of the information write-in/read-out system, electrical reading scheme is a prerequisite for any skyrmion-based information storage and computing device applications. Some methods were purposed to read and detect skyrmions in device applications, such as by using the magnetic tunnel junction (MTJ) or reading skyrmion-induced Hall voltages. In this section, we first review recent progress on the real-space imaging/reading of skyrmions in laboratories.

The microscopy imaging of magnetic skyrmions in real space is very important for both theoretical and applied reasons. From the theoretical point of view, the imaging of the in-plane and out-of-plane spin textures in real space can directly identify the topological nature of a skyrmion. For example, one may calculate the topological charge based on the detailed in-plane spin texture of a skyrmion, including the vorticity number $Q_v$ and helicity number $Q_h$. Moreover, some of techniques offer sub-ns temporal resolution along with real-space magnetization imaging, which plays an important role to understand the ultrafast dynamic properties of skyrmions, e.g. excitation modes.

The real-space imaging of a 2D skyrmion lattice, which consists of many Bloch-type skyrmions, was first realized in a thin film of $Fe_{0.5}Co_{0.5}Si$ by using the LTEM in 2010 [102]. Later, skyrmion lattices in thin films of FeGe [41, 103, 220-222], $Cu_2OSeO_3$ [109, 192, 234], MnSi [223, 235], $Mn_{1-x}Fe_xGe$ [224], and Fe-Gd alloy [225] were also observed by using the LTEM as discussed in Sec. 1.2. The LTEM is a particularly useful tool for observing the magnetization lying in the film plane, e.g. Bloch-type skyrmions, as the Lorentz interaction is non-vanishing when the magnetic induction component is perpendicular to the electron beam, which will deflect the electron beam [236]. Although typical Frensel-mode LTEM measurement cannot directly observe Néel-type magnetic configurations at zone-axis mode due to the cancellation of magnetic inductions, the observation of Néel-type magnetic domains and skyrmions has also been demonstrated using LTEM by tilting sample plane against beam direction or using off-focused images [237, 238]. More recently, as noted in Sec. 1.2. aberration-corrected DPC-STEM imaging technique has also been introduced [111, 239, 240] and used to directly image magnetic textures including skyrmions [111, 113].

The SP-STM can also be used to measure nanometer-scale magnetic textures (see Table 2). In 2013, Romming et al. observed nanoscale isolated skyrmions in a PdFe bilayer on Ir(111) by using SP-STM [106]. The field-dependent size and shape of single isolated skyrmions were then also identified by using the SP-STM [226]. The magnetic properties of nanoscale isolated skyrmions discovered in PdFe/Ir bilayers were also experimentally and theoretically investigated by Leonov et al. in 2016 [230]. In 2017, Hsu et al. further realized and observed the electric-field-induced switching of single isolated skyrmions in a Fe triple layer on Ir(111) using the SP-STM [195].

Recently years, various microscopy methods using X-ray radiation were used to as a powerful tool to image ultra-small spin textures and ultra-fast spin dynamics (see Table 2). In 2014, Li et al. observed the imprinted skyrmions by using the element-specific XMCD and PEEM measurements [174]. In 2016, Moreau-Luchaire et al. realized additive DM interaction in Ir/Co/Pt multilayers at room temperature, where isolated skyrmions with dimeters below 100 nm were directly observed by using the STXM [29]. In the same year, Boulle et al. observed room-temperature skyrmions in sputtered ultrathin Pt/Co/MgO multilayers by using the photoemission electron microscopy combined with XMCD-PEEM [120], which identified the Néel-type nature of the skyrmion chirality. Woo et al. also observed room-temperature skyrmions and revealed their current-driven dynamics in Pt/Co/Ta and Pt/CoFeB/MgO multilayers by using the high-resolution STXM [33, 121]. More recently, Woo et al. directly imaged ferrimagnetic skyrmions in GdFeCo films [227] and observed the writing and deleting dynamics of a ferrimagnetic skyrmion by using the time-resolved X-ray microscopy [36].

Indeed, skyrmion textures can also be directly observed by other methods. For example, by using the SPLEEM, Chen et al. imaged magnetic skyrmions in in-situ grown Fe/Ni bilayers in 2015 [125]. Moreover, X-ray holography offers a unique drift-free diffractive magnetic imaging characteristic with sub-10 nm spatial resolution (see Table 2), and was used to demonstrate ultra-small skyrmions and their current-driven dynamics [35, 241]. Recently, mainly inspired by easy accessibility, a number of experimental studies reported the direct observation of skyrmions by using the polar magneto-optical Kerr effect (MOKE) microscopy [116, 148, 160, 197, 228]. For example, the first observation of room temperature skyrmions and the creation of skyrmions from stripe domain walls was directly imaged by using the MOKE microscopy [116]. The spatial resolution of MOKE microscopy is rather limited by the wavelength of the light used for the observation, which usually lead to a resolution of a micrometer (see Table 2).

**Table 2. List of the representative imaging techniques for the observation of magnetic skyrmions**

| Methods | Spatial resolution | Advantages | Best used for | Reference |
|---|---|---|---|---|
| LTEM | ~ 1 nm | High resolution<br>Easy access | Thin chiral magnets with nanoscale Bloch-type skyrmions | [102, 109, 220-225] |
| DPC-STEM | ~ 1 nm | High resolution<br>No post imaging process | Thin chiral magnets with nanoscale Néel/Bloch-type skyrmions | [111, 113] |
| SP-STM | Single atom | Ultrahigh resolution down to atomic scale | Crystalline chiral magnets with extremely small skyrmions | [106, 195, 226] |
| XMCD-PEEM | ~ 25 nm | In-plane spin resolution | Surface magnetic skyrmions | [120, 174] |
| STXM | ~ 25 nm | Easy magnetic/electric excitation<br>Dynamics imaging (~70 ps) | Skyrmions in buried magnetic multilayers accompanying excitations | [29, 36, 227] |
| SPLEEM | ~ 10 nm | In-plane spin resolution<br>High surface sensitivity | In-situ deposited surface magnetic skyrmion | [125] |
| X-ray holography | ~ 10 nm | Drift-free technique<br>Dynamics imaging (~70 ps) | Nanoscale skyrmions in buried magnetic multilayers | [35] |

| MOKE | ~ 1 μm | Easy access and custom modification | >1 μm skyrmions | [116, 148, 160, 197, 228] |
|------|--------|--------------------------------------|------------------|---------------------------|

LTEM, Lorentz transmission electron microscopy;

DPC-STEM, differential phase contrast scanning transmission electron microscopy;

SP-STM, spin-polarized scanning tunneling microscopy;

XMCD-PEEM, photoemission electron microscopy combined with X-ray magnetic circular dichroism;

STXM, scanning transmission X-ray microscopy;

SPLEEM, spin polarized low-energy electron microscopy;

MOKE, magneto-optical Kerr effect microscopy.

## 3.2. Electrical Reading 1: Topological Hall Resistivity Measurements

The realization of reading skyrmions in an electrical manner is a crucial step toward the device applications based on skyrmions. A promising method to read skyrmions electrically is by harnessing the topological Hall effect (THE) of magnetic skyrmions [162, 242-244]. The THE arises from the Berry phase in a magnet with smoothly varying magnetization [245-247], which can occur due to the Berry curvature induced by magnetic skyrmions [162, 244], because conduction electrons must adiabatically obey the unique spin textures of magnetic skyrmions when the *s-d* coupling is strong [2]. Namely, a prominent feature of a skyrmion is that it produces emergent magnetic field, which originates from the solid angle subtended by the spins called scalar spin chirality. The topological property of skyrmion guarantees that the total flux generated by one skyrmion is one flux quantum, $h/e$. The size of a skyrmion is ~ 1 nm for atomic Fe layer on Ir(111) surface [8], ~ 3 nm for MnGe [244], ~ 18 nm for MnSi [243], and ~ 70 nm for FeGe [103]. The corresponding emergent magnetic field is ~ 4000 T, 1100 T, 28 T, and 1 T, respectively. In fact, the THE is a manifestation of emergent magnetic field, where the Hall effect occurs in the presence of skyrmions [107, 242-244]. Moreover, the quantized THE has been theoretically proposed [248].

In 2009, Neubauer et al. [242] experimentally studied the THE of the skyrmion crystal phase in MnSi, where a district anomalous contribution to the Hall effect in the skyrmion crystal phases was identified [see Fig. 10(a)]. Such a topological Hall contribution has an opposite sign to the normal Hall effect, and its prefactor is consistent with the skyrmion density in a quantitative manner. In 2012, Schulz et al. experimentally measured the temperature-dependent the Hall resistivity in MnSi under different strengths of applied electric current [107], where the depinning and motion of skyrmions were confirmed by direct observation of the emergent electric field of skyrmions. In the same year, magnetic skyrmions in epitaxial B20 FeGe(111) thin films was also identified by measuring the topological Hall resistivity due to the skyrmion textures [162].

In 2014, Yokouchi et al. studied the stability of skyrmions in epitaxial thin films of $Mn_{1-x}Fe_xSi$ with various thicknesses and compositions by measuring the topological Hall resistivity as functions of temperature and applied field [249]. The topological and thermodynamic stability of skyrmions in MnSi was also studied by measuring topological Hall resistivity [250]. In 2016, Matsuno et al. studied the interface DM interaction in epitaxial bilayers consisting of ferromagnetic $SrRuO_3$ and paramagnetic $SrIrO_3$ by measuring the THE [251], which demonstrated that the skyrmion phase is stabilized by interface-induced DM interaction. In 2017, Liu et al. demonstrated the creation and annihilation of skyrmion-induced THE in Mn-doped $Bi_2Te_3$ topological insulator films [252].

Although the electrical measurements induced by THE were mostly conducted with crystalline materials where the existence of skyrmion lattice provides large collective electrical signal, recent reports by Maccariello et al. [253] [see Fig. 10(b)] and Zeissler et al. [254] [see Fig. 10(c)] demonstrated the electrical Hall measurements of individual room temperature skyrmion in sputter-grown films and nanostructures. However, the two demonstrations report large difference in the effective magnitude of topological Hall contribution from a single room-temperature skyrmion, which may need further investigations. In 2019, Shao et al. also experimentally observed the THE at above room temperature in a bilayer heterostructure composed of a magnetic insulator (thulium iron garnet, $Tm_3Fe_5O_{12}$) in contact with a metal (Pt).

On the other hand, it is noteworthy that, in 2016, Hamamoto et al. theoretically proposed a method to detect the skyrmion position in a pure electrical way by measuring the Hall conductance in a constricted geometry [255], where the Hall conductance is found to have a peak when a skyrmion is located at the lead position, while it reduces when a skyrmion is away from the lead.

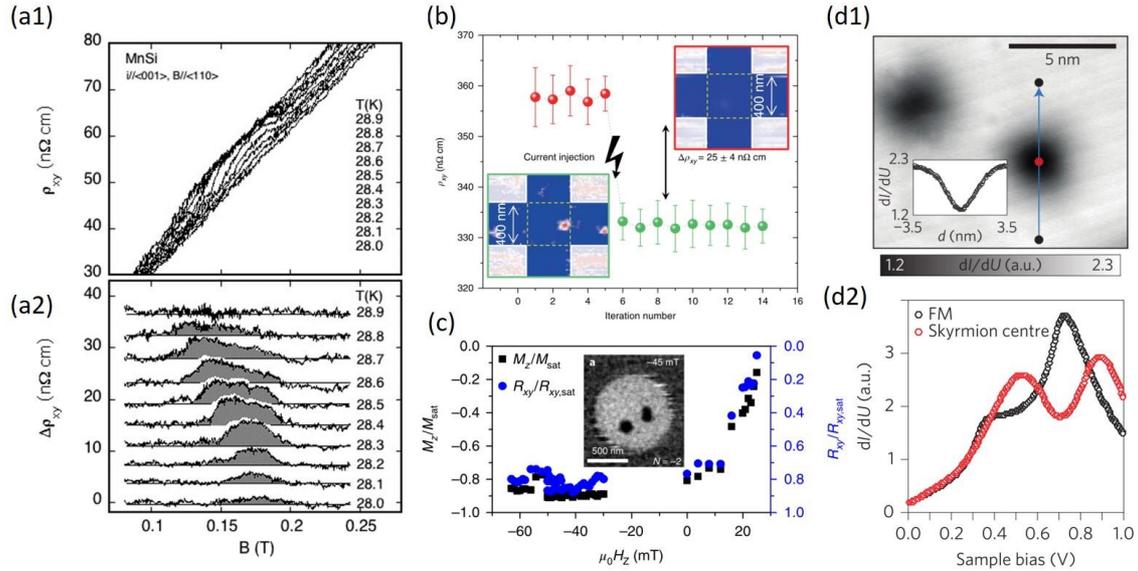

Figure 10. Electrical reading of magnetic skyrmions by using the THE and MTJ. (a) Topological Hall effect in the A phase of MnSi. (a1) Hall resistivity $\rho_{xy}$ near Tc in the temperature and field range of the A phase. (a2) Additional Hall contribution $\rho_{xy}$ in the A phase. Data are shifted vertically for better visibility. Reprinted with permission from [242]. Copyright © 2009 The American Physical Society. (b) Topological Hall effect in skyrmion phase of MnSi. (b) Variation in Hall due to a single skyrmion. Hall resistivity shows a change of 25 n$\Omega$ cm associated with the single-skyrmion formation. The insets are MFM images. Reprinted with permission from [253]. Copyright © 2018 Macmillan Publishers Limited, part of Springer Nature. (c) The normalized Hall resistance $R_{xy}/R_{xy,sat}$ and the extracted $M_z/M_{sat}$. The inset is the XMCD image of two skyrmions at -45 mT. Reprinted with permission from [254]. Copyright © 2018 Macmillan Publishers Limited, part of Springer Nature. (d) Electrical detection of magnetic skyrmions by tunnelling non-collinear magnetoresistance. (d1) The signal change caused by the non-collinear magnetoresistance dI/dU map of two skyrmions in PdFe/Ir(111); the inset presents a profile along the arrow. (d2) The dI/dU tunnel spectra in the centre of a skyrmion (red) and outside the skyrmion in the FM background (black). a.u., arbitrary units. Reprinted with permission from [256]. Copyright © 2015 Macmillan Publishers Limited.

## 3.3. Electrical Reading 2: Magnetic Tunnel Junction (MTJ)

Recently, theoretical and experimental studies suggest the efficient large on/off ratio detection of skyrmions by measuring the magnetoresistance [256-259], i.e. by using and embedding MTJ sensors in skyrmion-based devices and circuits.

In 2015, Du et al. experimentally identified the formation of clusters of individual skyrmions by measuring the magnetoresistance in MnSi nanowires [229, 257], where the number of skyrmions was revealed by quantized jumps in the magnetoresistance curves. In the same year, Hanneken et al. [256] proposed a pure electrical reading scheme of nanoscale skyrmions by measuring the tunnelling non-collinear magnetoresistance [see Fig. 10(d)]. The electric conductance signal due to the non-collinear magnetoresistance is sensitive to the local magnetic environment, and can directly detect and distinguish the collinear and non-collinear magnetic states without using a magnetic electrode [256, 259]. Such a mechanism is different from the well-known giant magnetoresistance (GMR) [260], and tunnel magnetoresistance (TMR) effect [261], which are dependent of local magnetization directions in two magnetic layers. On the other hand, the electric reading of single isolated skyrmions in a current-perpendicular-to-plane geometry was also studied by Crum et al. from first principles [262], which is based on the effect of tunnelling spin-mixing magnetoresistance.

In 2017, Tomasello et al. theoretically proposed a protocol for the electrical reading of a magnetic skyrmion by measuring the change of the TMR signal via a point-contact MTJ in a three-terminal device [258]. Similar method can also be used to read other magnetic solitons, such as magnetic bubbles, in racetrack-type devices [263]. Most recently, numerical simulations suggested that skyrmions can be directly created and detected in a MTJ with DM interaction in a pure electrical manner [264], which shows the possibility of designing the skyrmion-MTJ-based multibit storage and artificial neural network computation. In 2019, Penthorn et al. found that a single skyrmion with a diameter smaller than 100 nm can lead to a change in MTJ resistance of almost 10% [265], which is enough for electrical detection. In the same year, Kasai et al. also experimentally realized the electrical detection of skyrmions in MTJ [266], where the skyrmion diameter is about 200 nm.

## 4. Processing Skyrmions

### 4.1. Current-Driven Dynamics of Magnetic Skyrmions

As briefly mentioned in Sec. 1.2, Jonietz et al. [78] observed the rotation of skyrmions arising from the interplay of current-induced spin-torque effects and thermal gradients. They found that the current density required to create observable STTs in skyrmion lattice phase of MnSi exceeds an ultralow threshold of $\sim 10^6$ A m$^{-2}$, which is over five orders of magnitude smaller than those typically applied in experimental studies on current-driven magnetization dynamics in nanostructures ($\sim 10^{11}$ A m$^{-2}$). These results indicate that the current-driven motion of magnetic skyrmions could be of low-energy consumption. Therefore, the spin-polarized current serves as an effective method for driving and processing skyrmions. However, it should be noted here that the skyrmion velocity is proportional to the driving current density, and practical skyrmion-based applications may require a skyrmion speed much larger than that induced by the very low depinning current density. At the same time, the energy consumption of skyrmion-based application will increase with increasing magnitude of the current.

In 2011, Everschor *et al.* [180] theoretically proposed that skyrmions can be driven by the spin-polarized current via STTs. The translational mode and rotational mode can be induced by STTs. Using the method of Thiele [267], they projected the magnetization dynamics equation, i.e. the Landau-Lifshitz-Gilbert (LLG) equation onto the translational mode to derive an effective equation of motion – Thiele equation,

$$\boldsymbol{G} \times (\boldsymbol{v_s} - \boldsymbol{v_d}) + \boldsymbol{\mathcal{D}}(\beta \boldsymbol{v_s} - \alpha \boldsymbol{v_d}) = \boldsymbol{0}, \tag{15}$$

where $\boldsymbol{G}$ is the gyrocoupling vector and $\boldsymbol{\mathcal{D}}$ is the dissipative tensor. $\boldsymbol{v_s}$ is the velocity of the conduction electrons and $\boldsymbol{v_d}$ is the drift velocity of skyrmions.

In 2011, Zang et al. [26] considered the additional damping term in the LLG equation and theoretically proposed the skyrmion Hall effect (SkHE), where the skyrmion shows a transverse velocity when it is driven by the spin current. In 2013, the current-induced creation and motion of skyrmions are numerically demonstrated by Iwasaki et al. [108, 132] and Sampaio et al. [133] [see Fig. 11(a)]. Iwasaki et al. investigated the current-induced motion of skyrmions in the presence of geometrical boundaries [108, 132]. In a channel with a finite width, the transverse confinement results in steady-state characteristics of the skyrmion velocity as a function of current density, which are similar to those of domain walls in ferromagnets. Sampaio et al. [133] considered two cases. One is the case of spin-polarized current injected under a current-in-plane geometry. The skyrmion is driven by the adiabatic and non-adiabatic STTs,

$$\boldsymbol{\tau}_{\text{adiab}} = u\boldsymbol{m} \times \left(\frac{\partial \boldsymbol{m}}{\partial x} \times \boldsymbol{m}\right), \boldsymbol{\tau}_{\text{non-adiab}} = \beta u \left(\boldsymbol{m} \times \frac{\partial \boldsymbol{m}}{\partial x}\right), \tag{16}$$

where $u = \left|\frac{\gamma_0 \hbar}{\mu_0 e}\right| \frac{jP}{2M_S}$ with the reduced Planck constant $\hbar$, the electron charge $e$, the applied current density $j$, the spin polarization rate $P$, and the saturation magnetization $M_S$. is the electron velocity and $\beta$ is the non-adiabaticity factor. The other is the case driven by a vertical spin current, which can be obtained either by using a MTJ or by utilizing the spin Hall effect. The skyrmion is driven by the induced in-plane and out-of-plane torques,

$$\tau_{\text{IP}} = \frac{u}{t}\boldsymbol{m} \times (\boldsymbol{m}_{\text{p}} \times \boldsymbol{m}), \tau_{\text{OOP}} = -\xi\frac{u}{t}(\boldsymbol{m} \times \boldsymbol{m}_{\text{p}}), \tag{17}$$

where $t$ is the film thickness and $\xi$ is the amplitude of the out-of-plane torque relative to the in-plane one and $\boldsymbol{m}_{\text{p}}$ is the current polarization vector. The numerical results obtained by Sampaio et al. [133] demonstrated that the efficiency of the vertical spin current to drive skyrmions is much higher than that of the in-plane current.

In 2016, Woo et al. [121] experimentally demonstrated the stabilization of magnetic skyrmions and their current-driven motion in thin transition metal ferromagnets at room temperature, where a train of individual skyrmions can be driven into motion at speeds exceeding 100 m s$^{-1}$ by current pulses ($5 \times 10^{11}$ A m$^{-2}$) in the Pt/CoFeB/MgO multilayers track. In 2017, Jiang et al. [218] and Litzius et al. [219] directly observed the SkHE [see Fig. 11(b)] in room-temperature experiments where skyrmions are driven by the current-induced SOTs. Like the charged particles in the conventional Hall effect, current-driven skyrmions acquire a transverse velocity component. As shown in Fig. 11(c), Jiang et al. observed a linear dependence of the skyrmion Hall angle on the driving current density, which is possibly resulted by the pinning of skyrmions. They also changed the sign of the topological charge $Q$ and the electric current, and thus, a strong similarity between the conventional Hall effect of the electronic charge was found. Litzius et al. also experimentally observed the dependency of the skyrmion Hall angle on the driving current density, while they suggested that this dependency is induced by the additional effect of the skyrmion deformation as well as the effect of the field-like SOT.

In 2017, Legrand et al. [32] investigated the nucleation and spin-torque-induced motion of sub-100 nm skyrmions in Pt/Co/Ir trilayers, which enable additive DM interactions at the Pt/Co and Co/Ir interfaces [29]. In particular, they showed that such small compact skyrmions can be nucleated by applying a uniform current directly into tracks and subsequently be driven into motion via SOTs.

In 2016, theoretical studies [155, 268] have predicted enhanced current-driven behaviors of antiferromagnetic skyrmions due to the elimination of SkHE. In 2018, Woo et al. [36] experimentally investigated the skyrmion in ferrimagnetic GdFeCo multilayers, where the stabilization of ferrimagnetic skyrmions and their current-driven dynamics were demonstrated. The distinctive nature is that the SkHE is not totally eliminated in the ferrimagnetic system since the magnetization are different between two antiferromagnetically coupled underlying sub-lattices. Indeed, they demonstrated that the current-driven ferrimagnetic skyrmion shows a small, but non-zero SkHE. They further confirmed that ferrimagnetic skyrmions can move at a velocity of ~50 m s⁻¹ with reduced skyrmion Hall angle, |θ_SkHE| ~20°. In 2019, Hirata et al. [269] experimentally demonstrated that the SkHE can be totally vanished at the angular momentum compensation temperature of a ferrimagnet [see Fig. 11(d)], i.e. when the ferrimagnet becomes antiferromagnet.

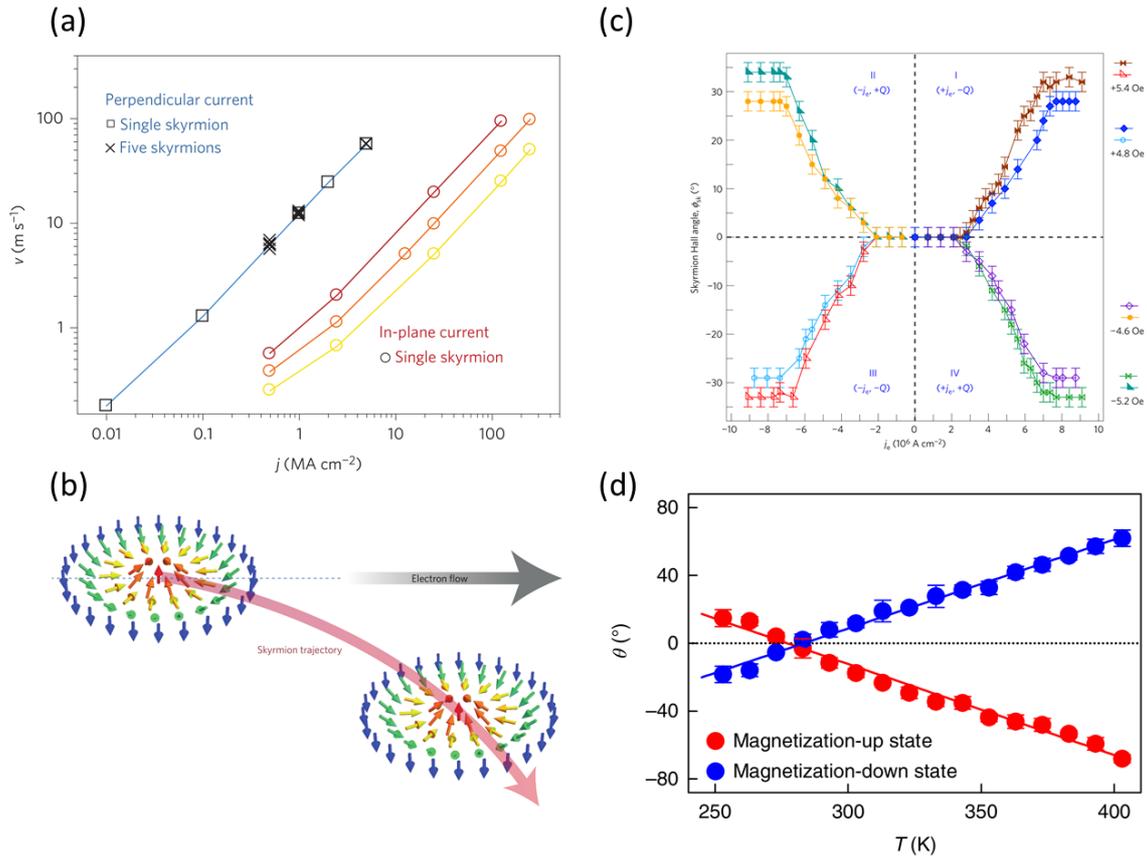

Figure 11. (a) Skyrmion velocity v as a function of current density j for in-plane currents with different values of the non-adiabaticity parameter β (0.15, 0.30 and 0.60 in yellow, orange and brown lines and circles, respectively) and for vertical currents (blue line, squares for isolated skyrmion). Reprinted with permission from [133]. (b) Schematic of the skyrmion Hall effect. The spin textures of skyrmions are indicated by the arrows. Reprinted with permission from [218]. (c) Phase diagram of the skyrmion Hall angle as a function of current density/sign of topological charge obtained by tracking the motion of several tens of skyrmions. Reprinted with permission from [270]. (d) Current-driven elongation of magnetic bubble in GdFeCo/Pt films as a function of temperature. Reprinted with permission from [269].

**4.2. Information Storage: Racetrack Memory and Skyrmion-MTJ**

In 2013, Fert et al. [1] proposed that the skyrmion can be used to build the racetrack memory, which is based on the design of domain wall-based racetrack memory [271]. In the skyrmion-based racetrack memory, the information can be coded by the presence and absence of skyrmions, as shown in Fig. 12(a). In 2014, Tomasello et al. [272] showed the technological advantages and limitations of the manipulation of Bloch-type and Néel-type skyrmions by spin currents generated in the ferromagnetic layer or in the heavy-metal substrate arising from the spin Hall effect. They found that the Néel-type skyrmion driven by the spin torques due to the spin Hall effect is a promising strategy for technological implementation of next-generation of skyrmion racetrack memories. In 2017, Yu et al. experimentally demonstrated a room-temperature skyrmion shift device [273], which could serve as a basis for building a fully functional skyrmion-based racetrack memory.

In 2014, Zhou et al. numerically demonstrated a current-driven reversible conversion between a skyrmion and a domain-wall pair in a junction geometry [182]. The information encoded in domain walls can be transformed into skyrmions, and then read out by transforming skyrmions back to domain walls after a functional control of the skyrmions. Such a hybrid device has the potential to outperform domain-wall racetrack memory because that it combined advantages of domain walls and skyrmions for spintronic applications.

In 2015, Zhang et al. [274] numerically investigated the effects of skyrmion-skyrmion and skyrmion-edge repulsions on the feasibility of skyrmion-based racetrack memory. They suggested that the practicable spacing between consecutive skyrmionic bits is determined by the DM interaction helix length, i.e. $4\pi A/|D|$, where $A$ is the exchange constant and $D$ is the DM interaction constant. Further, they demonstrated that by fabricating a notched tip at the end of the nanotrack is an effective and simple method to avoid the clogging of skyrmionic bits, which also enables the elimination of useless skyrmionic bits beyond the reading element.

As discussed in Sec. 4.1, when the skyrmion is driven into motion by a spin current, the trajectory of skyrmion bends away from the driving current direction due to the topological Magnus force [108, 155, 275], which is always perpendicular to the velocity. Such a SkHE may create an obstacle for the transportation of skyrmions. Consequently, the skyrmions may be destroyed by touching the nanotrack edges. In 2016, Zhang et al. [276] theoretically proposed a antiferromagnetically exchange-coupled bilayer system, i.e. a synthetic antiferromagnetic bilayer nanotrack, in which the SkHE can be suppressed as the Magnus forces in the top and bottom layers are exactly cancelled. This system serves as a promising platform where skyrmions can move in a perfectly straight trajectory even at an ultra-fast processing speed. In 2017, Tomasello et al. [277] investigated the performance of synthetic antiferromagnetic racetrack memory. They pointed out that two adjacent racetracks can be more closer as compared with the with single heavy metal/ferromagnet bilayer since the dipole interactions of the two antiferromagnetically coupled layers cancel each other and the disturbance is thus reduced significantly.

In 2016, Kang et al. [278] also proposed a complementary skyrmion racetrack memory structure, in which the skyrmions can be selectively driven into two different nanotracks by using a voltage-controlled Y-junction. In such proposed structure, both data bits "0" or "1" are represented with the presence of a skyrmion, therefore improving the data robustness and clock synchronization. In 2018, a skyrmion-based multilevel device with tunneling magnetoresistance was proposed [264], in which a MTJ with stable intermediate states can be realized based on the skyrmionic states in the free layer [see Fig. 12(b)]. This design may also facilitate the electrical detection of skyrmions. Most recently, the deterministic electrical switching from a ferromagnetic state to a skyrmion spin texture in the free layer of a MTJ was experimentally realized by Penthorn et al. [265]. Kasai et al. also experimentally realized the electrical detection of skyrmions in MTJ [266]. Although fully functional skyrmion-based racetrack memory has not been experimentally demonstrated yet, integrating above mentioned findings in a single device scheme may enable the realization of such skyrmion-based information storage devices.

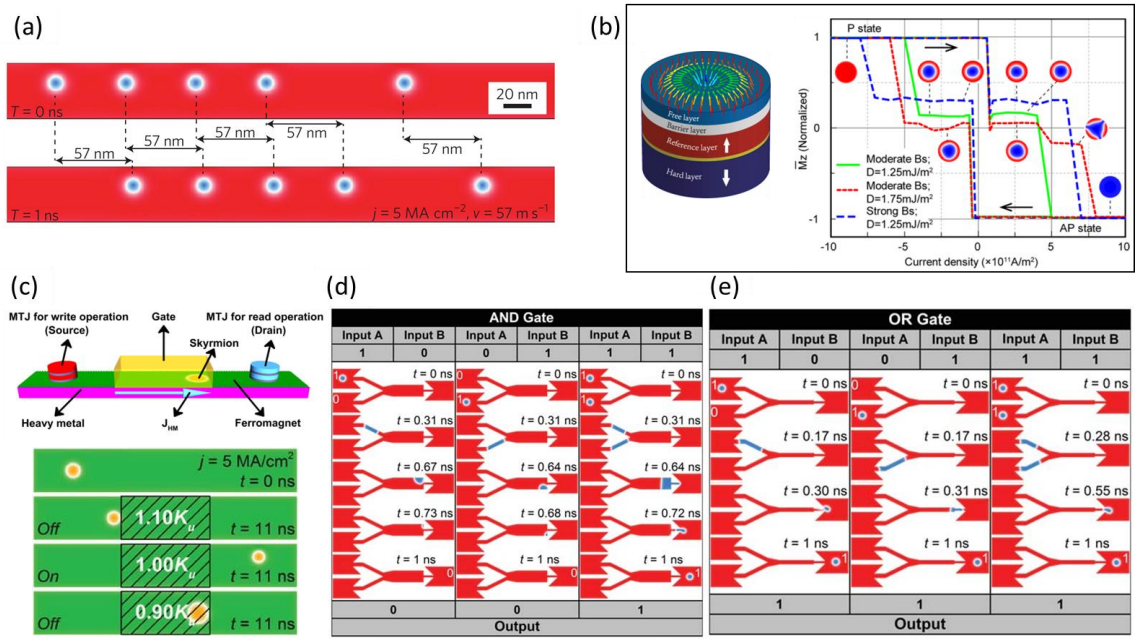

Figure 12. (a) The illustration of skyrmionic racetrack memory, in which the skyrmions exhibit the same velocity. The colour scale shows the out-of-plane component of magnetization, $m_z$. Reprinted with permission from [1]. (b) Illustration of a skyrmion-based MTJ and Magnetization−current hysteresis loops obtained by simulating the switching of $R = 60$ nm MTJs with different $D$ and stray field $B_S$. Reprinted with permission from [264]. (c) The top-view of the nanotracks under different spin current density j as well as different voltage-controlled perpendicular magnetic anisotropy $K_{uv}$. Initial state: both the electric field and spin current are turned off; the skyrmion keeps its position on the left side of the nanotrack. Off state: both the electric field and spin current are turned on. The spin current drives the skyrmion moving toward the right, while the electric field, which results in the change of PMA in the voltage-gated region, leads to the termination of the skyrmion when it approaches the voltage-gated region. On state: the electric field is turned off but the spin current is turned on. The skyrmion driven by the spin current passes the voltage-gated region and reaches the right side of the nanotrack. Reprinted with permission from [141]. (d) Skyrmion logical AND operation. The skyrmion represents logical 1, and the ferromagnetic ground state represents logical 0. Left panel, the basic operation of AND gate 1 + 0 = 0. Middle panel, the basic operation of the AND gate 0 + 1 = 0. Right panel, the basic operation of the AND gate 1 + 1 = 1. (e) Skyrmion logical OR operation. The skyrmion represents logical 1, and the ferromagnetic ground state represents logical 0. Left panel, the basic operation of OR gate 1 + 0 = 1. Middle panel, the basic operation of the OR gate 0 + 1 = 1. Right panel, the basic operation of the OR gate 1 + 1 = 1. (d) and (e) are reprinted with permission from [140].

## 4.3. Information Computing: Transistor-Like Devices and Skyrmionic Logic

In 2015, Zhang et al. [141] proposed a skyrmion-based transistor-like functional device [see Fig. 12(c)], where a gate voltage can be used to switch on/off a circuit. The PMA in the gate region is locally controlled by an applied electric field due to the charge accumulations. For the ON state: the spin current is turned on but the electric field is turned off. The skyrmion driven by the spin current passes the voltage-gated region. For the OFF state: both the electric field and the spin current are turned on; the electric field changes the PMA in the voltage-gated region and creates an energy barrier, leading to the termination of the skyrmion when it approaches the voltage-gated region. Zhang et al. numerically demonstrated that the working conditions can be controlled by tuning the amplitude of applied electric field and spin current, while proving the scalability of this transistor-like device. In the same year, a similar idea on the control of skyrmion dynamics by a gate voltage was also proposed by Upadhyaya et al. [142], which also suggests the transistor-like function of skyrmions. In 2017, Xia et al. [279] numerically demonstrated the skyrmion-based transistor-like functional device also can be driven by spin waves and the working conditions can be adjusted by the amplitude and frequency of spin waves, as well as the applied voltage on the gated region. Later in 2018, Zhao et al [280] also numerically demonstrated the antiferromagnetic skyrmion transistor-like device based on strain manipulation.

In 2015, Zhang et al. [140] theoretically proposed that skyrmions can be reproduced and merged in junction geometries based on the reversible conversion between skyrmions and domain walls [182]. They demonstrated that the logic AND [see Fig. 12(d) and OR [see Fig. 12(e)] operations can be realized based on the duplication and merging of skyrmions. In their proposed skyrmionic logic devices, binary digit "0" corresponds to the absence of a skyrmion and binary digit "1" corresponds to the presence of a skyrmion. There are two input branches and one output branch in the device. The logic AND operations, "0" + "0" = "0", "0" + "1" = "0", "0" + "1" = "0", "1" + "1" = "1", can be implemented in the designed geometry. For example, "1" + "0" means that there is one skyrmion in input A and no skyrmion in input B. Under the driving force provided by the spin current, the skyrmion in branch A moves toward the output side. The skyrmion is first converted into a domain wall pair and the domain pair fail to convert into skyrmion in the wide Y-junction. Then, no skyrmion exists in the output branch. Thus, the logic operation "1" + "0" = "0" is thus realized. Similarly to the AND operation, the OR operation, "0" + "0" = "0", "0" + "1" = "1", "0" + "1" = "1", and "1" + "1" = "1", can also be implemented in a slightly modified geometry. For example, in the OR operation, "1" + "0" means that there is one skyrmion in input A and no skyrmion in input B. Under the driving force provided by the spin current, the skyrmion in branch A moves toward the output side. The skyrmion is first converted into a domain wall pair and the domain pair is then converted into skyrmion in the narrow Y-junction. Therefore, there is one skyrmion in the output branch finally, and the operation "1" + "0" = "1" is implemented.

Soon after the numerical demonstration of the first prototype of skyrmionic logic computing device, several works on skyrmion logic emerged. For example, in 2015, Zhang et al. [281] designed the skyrmion-based NIMP, XOR, and IMP gates. In 2016, Xing et al. [282] numerically demonstrated the NAND and NOR gates by using both domain walls and skyrmions. In 2018, Luo et al. [283] numerically demonstrated logic functions including

AND, OR, NOT, NAND, NOR, XOR, and XNOR in the ferromagnetic nanotrack by virtue of various effects including SOT, SkHE, skyrmion-edge repulsions, and skyrmion-skyrmion collision. However, it should be noted that the skyrmion-based Boolean logic operations have not yet been demonstrated in an experiment. The complex structures and their co-existences of the already proposed skyrmion logic gates as well as the precise control of the dynamic behaviors of skyrmions may be difficult to be realized in nanoscale dimensions at the current stage.

## 4.4. Bio-Inspired Computing: Skyrmions for Neuromorphic Devices

In 2017, Huang et al. proposed that the skyrmion can be used to build a skyrmion-based artificial synapse device for neuromorphic computing. They numerically demonstrated the short-term plasticity and long-term potentiation functions based on skyrmions in a nanotrack. The synaptic weight of the proposed device can be strengthened/weakened by positive/negative stimuli, mimicking the potentiation/depression process of a biological synapse. Also, the resolution of the synaptic weight can be adjusted based on the nanotrack width and the skyrmion size. The merits of using skyrmions for such artificial synaptic device include several aspects. For example, the skyrmion structure can be regarded as a nanoscale rigid object due to its quasi-particle-like nature. Based on this feature, many nanoscale skyrmions can accumulate in a given device and controlled by external stimuli, of which the properties are approximately analogous to those of a biological synapse and could be used to build nanoscale computing devices with good variability and scalability [3, 106]. Moreover, the extremely low-threshold for skyrmion excitation [1, 78, 107, 108] may promise the reduced power consumption of skyrmion synapse-based neuromorphic computing, which is highly desired for such device to be used for massive parallel computations. Later in 2017, Li et al. [145] presented a magnetic skyrmion-based artificial neuron device concept based on current-induced skyrmion motion in a nanotrack. The neuronal activity of a biologic neuron was realized with the tunable current-driven skyrmion motion dynamics of a single skyrmion. In 2018, Chen et al. [284] also proposed a magnetic skyrmion-based device to emulate the core functionality of neurons and synapses for an all-spin spiking deep neural network. The synaptic weight can be adjusted by the number of skyrmions under the read MTJs, and the resolution can be improved by having multiple branches with various conductance ranges. Recently, utilizing current-induced skyrmion generation, motion, deletion and detection together in a single device scheme, Song et al. experimentally demonstrated the basic operations of a magnetic skyrmion-based artificial synapse including *potentiation* and *depression*, and further demonstrated artificial skyrmion synapse-based neuromorphic pattern recognition computing using simulations [144].

Besides, in 2018, Prychynenko et al. [147] proposed the skyrmion-based reservoir computing applications, in which a single skyrmion is embedded in a ferromagnetic ribbon. The reservoir computing was realized via a two-terminal device with nonlinear voltage characteristics originating from magnetoresistive effects. Most recently in 2019, Zázvorka et al. experimentally demonstrated a thermal skyrmion diffusion-based signal reshuffling device [160]. In this work, using both experiment and simulation, Zázvorka et al. uncovered the dynamics of thermally activated skyrmion diffusions using low-pinning multilayer materials, and further analysis revealed the possibility of using skyrmions for an efficient future probabilistic computing device [160].

## 5. Summary and Outlook

### 5.1. Potential Novel Materials: Antiferromagnet, Synthetic Antiferromagnet, Ferrimagnet, Frustrated Magnet and 2D van der Waals Magnet

In this section, we give an outlook on magnetic skyrmions in terms of potential novel materials and other possible topological spin textures. From the viewpoint of possible skyrmion-hosting materials, many previous works have focused on the ferromagnet, which is the most commonly known and important class of magnetic materials that has been employed in commercial products, such as modern hard disk drives. There are also two other technologically promising classes of magnetic materials, namely, the ferrimagnets and antiferromagnets, which can be used to host skyrmions. For example, several theoretical studies have demonstrated that skyrmion textures can be stabilized and manipulated in antiferromagnets [155, 268, 285-287].

For the last few years, antiferromagnetic spintronics has received much attention from the magnetism community as the antiferromagnetic materials have several intrinsic properties that may lead to a better performance of spin dynamics suitable for practical applications [288, 289]. Since 2016, several theoretical works have suggested that the magnetic skyrmions in antiferromagnets may have better mobility in compared to that in ferromagnets [155, 268, 290-293]. In ferromagnets, the magnetic skyrmion driven by an external force, such as the spin current, usually shows an undesired transverse shift, which may result in the destruction of skyrmion at sample edges. Such a phenomenon is referred to as the SkHE as mentioned in Sec. 4, which is attributed to the Magnus force induced by the topological nature of the skyrmion. In antiferromagnets, as shown in Fig. 13(a), a magnetic skyrmion can be regarded as two coupled sublattice skyrmions with opposite topological charges, which means that the Magnus forces acted on the two sublattice skyrmions can be exactly cancelled and therefore, leading to the straight motion of the antiferromagnetic skyrmion along the direction force direction [155, 268, 290]. Compared to the current-driven skyrmions in ferromagnets, the speed of antiferromagnetic skyrmions can also be significant improved [155, 268, 290]. Similar to the skyrmions in antiferromagnets, the bilayer skyrmions in the synthetic antiferromagnets [see Fig. 13(b)] also have a zero topological charge and show no SkHE when they are driven by external driving forces [276, 277, 294]. Theoretical works also suggest that the multilayer skyrmions in a synthetic antiferromagnetic multilayer packed with even constituent ferromagnetic layers are also immune from the SkHE [30]. However, the efficient electrical manipulation (in particular – detection) of antiferromagnetic spin textures with zero net magnetization remains as an important challenge for the practical application of antiferromagnetic materials for skyrmion-electronics.

On the other hand, ferrimagnets are also composed of two sublattices which favor antiparallel alignment of spins between each other, where the magnetic moments of the two sublattices are not equal and lead to a net magnetic moment. Therefore, the ferrimagnetic skyrmions [see Fig. 13(c)] should have intermediate properties between ferromagnetic and antiferromagnetic ones. Indeed, ferrimagnetic skyrmions have non-zero SkHE, but their SkHE is much reduced in compared to the ferromagnetic skyrmions [227]. Recent reports also demonstrated that compensated ferrimagnets can have zero SkHE [269] as well as largely improved current-driven velocity suggested by the current-driven motion of domain walls in compensated ferrimagnets [241]. Due to the absence or reduction of the SkHE and their inherent fast dynamics, both antiferromagnetic and ferrimagnetic skyrmions can reach a very high speed without being destroyed at sample edges on a racetrack-like device scheme. That is to say, the antiferromagnetic skyrmion can strictly move along the nanotrack without showing a transverse motion toward the edge. For the ferrimagnetic skyrmion, although it still shows a transverse motion (i.e., a shift) toward the edge, the skyrmion-edge interaction becomes very strong only when the skyrmion velocity is extremely large. Therefore, antiferromagnets, synthetic antiferromagnets, and ferrimagnets are considered as possible skyrmion-hosting materials, where the dynamic performance of skyrmions can be remarkably improved in compared with their counterparts in conventional ferromagnetic materials.

Recently, another type of magnetic materials with exotic and particular properties, i.e. the frustrated magnet, has been reported to host magnetic skyrmions. The frustration [295] in magnets can be raised from competing exchange interactions, and usually leads to complex magnetic phase diagrams. In 2012, Okubo et al. theoretically discovered that the skyrmion lattice can be stabilized in a frustrated magnet by an order-from-disorder mechanism [89]. In 2015, Leonov and Mostovoy theoretically investigated a rich phase diagram of an anisotropic frustrated magnet and properties of frustrated skyrmions with arbitrary vorticity and helicity [296]. Then, many exotic and unique properties of frustrated skyrmions have also been studied theoretically [42, 297-300]. For example, the energy of a skyrmion with a vorticity number of $Q_v = +1$ is identical to that of an antiskyrmion with a vorticity number of $Q_v = -1$. Namely, the skyrmion energy in frustrated magnets can be irrespective to the skyrmion helicity [301]. Also, a frustrated skyrmion may show coupled dynamics of its helicity and center of mass, which can result in the rotational skyrmion motion [301-303]. Indeed, the coupled dynamics of a frustrated skyrmion depends on many factors. In 2017, Zhang et al. [302] theoretically reported the current-induced helicity locking-unlocking transition phenomenon of a frustrated skyrmion, which suggests the dipolar interaction may also play an important role in the dynamics of skyrmions in the frustrated magnets, especially at low temperature.

Owing to the multiple degrees of freedom of skyrmions in frustrated magnets, it is anticipated that frustrated skyrmions can be used as versatile information carriers to perform information storage and computing. As shown in Fig. 13(d), in 2017, Hou et al. successfully observed various and spontaneous skyrmion bubbles in a frustrated kagome magnet $Fe_3Sn_2$ by using the LTEM at room temperature [129]. In 2019, Kurumaji et al. [304] experimentally found a Bloch-type skyrmion state in a frustrated centrosymmetric triangular-lattice magnet $Gd_2PdSi_3$, where a giant topological Hall response is observed, indicating the field-induced skyrmion phase.

Very recently, it has been reported that 2D van der Waals (vdW) crystals can also have long-range intrinsic ferromagnetism in few materials with strong magnetic anisotropy [305, 306], e.g. $Cr_2Ge_2Te_3$ or $CrI_3$, whereas such magnetic order is usually strongly suppressed by thermal fluctuations [see Fig. 13(e)], as predicted by the Mermin-Wagner theorem [307]. These studies on 2D magnets opened a whole new door toward 2D heterostructure-based spintronics, as a result, there appeared significant following interests and efforts revealing the possibility of such devices by demonstrating the magnetoresistance effect in 2D magnets [308], gate-tunable room-temperature 2D magnets [309], current-driven magnetization switching [310, 311], skyrmions in 2D magnets [312], and so on. Due to the broken inversion symmetry and expected large SOC (induced by heavy atomic component) of some representative 2D magnets, e.g. $Fe_3GeTe_2$, it was suggested that 2D magnets might permit the presence of DM interaction that could stabilize magnetic skyrmions [313, 314]. Recently, several experimental studies indeed demonstrated the stabilization of Néel-type magnetic skyrmions in a van der Waals ferromagnet such as $Fe_3GeTe_2$ [315-317] and $Cr_3Ge_2Te_6$ [318], which may excite large research efforts on 2D van der Waals magnet-based skyrmions.

Nevertheless, continuous studies on potential novel skyrmion-hosting materials could reveal more unexpected but remarkable properties and features of skyrmions that can be utilized to perform more efficient information processing in the future.

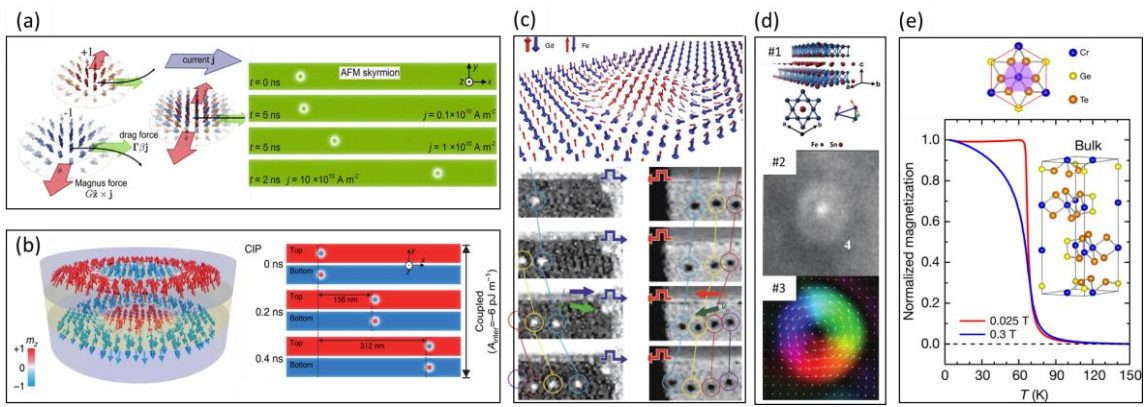

Figure 13. Skyrmions in potential novel materials. (a) The AFM Skyrmion is composed of two topological objects with opposite topological charge; hence, the Magnus force acts in opposite directions. The strong coupling between the sublattices leads to a perfect cancellation of the two opposing forces, and so, the AFM Skyrmion has no transverse motion. Reprinted with permission from [155] and [268]. (b) Illustration of a bilayer skyrmion in an AFM-coupled nanodisk and the current-induced motion of skyrmions in the top and bottom FM layers of an AFM-coupled bilayer nanotrack. Reprinted with permission from [276]. (c) Schematic of antiferromagnetically exchange-coupled ferrimagnetic skyrmion on a magnetic track as observed in GdFeCo films and sequential STXM images showing the responses of multiple ferrimagnetic after injecting unipolar current pulses. Reprinted with permission from [227]. (d) Skyrmionic bubbles in a frustrated kagome Fe3Sn2 magnet imaged using LTEM at 300 K. Reprinted with permission from [129]. (e) Temperature-dependent magnetization of the bulk crystal $Cr_2Ge_2Te_6$ measured by SQUID under fields of 0.025 T (red) and 0.3 T (blue). The inset shows the crystal structure of $Cr_2Ge_2Te_6$. Bulk $Cr_2Ge_2Te_6$ has a layered structure with interlayer vdW spacing of 3.4 Å. Reprinted with permission from [305].

## 5.2. Other Topological Spin Textures: Antiskyrmion, Skyrmionium, Biskyrmion, Meron, Antimeron, and Bimeron

During extensive studies on magnetic skyrmions since a decade ago, a number of derivative concepts on skyrmion-like objects were created and envisioned for practical applications. As reviewed in Sec. 1, the skyrmion structure and its topological charge are determined by both in-plane and out-of-plane spin configurations. That is to say, there could be a variety of different skyrmion-like objects due to the modification of either the in-plane or out-of-plane spin configuration in principle. For example, most studies have focused on Néel-type and Bloch-type skyrmions with topological charge of $|Q| = 1$, however, several recent theoretical and experimental reports have revealed that antiskyrmions [81, 88, 205, 300, 319-322] and skyrmioniums [58, 61, 71, 72, 209, 323] have special dynamic properties and can be employed as information carriers similar to conventional skyrmions.

In compared to skyrmions with $Q_v = +1$, the antiskyrmion has different in-plane spin textures, which leads to an opposite-sign vorticity of $Q_v = -1$ as well as largely different current-induced motion as reported in several recent studies [301-303, 314, 324]. In 2017, Nayak et al. [88] for the first time observed the antiskyrmions in tetragonal Heusler materials above room temperate, as shown in Fig. 14(a). Also, both skyrmions and antiskyrmions are stable or meta-stable solutions in frustrated magnets [301-303]. The skyrmion with $Q_v = +1$ and antiskyrmion with $Q_v = -1$ can be used to carry different binary information bits in a same spintronic device. Namely, the skyrmion stands for digital "1", and the antiskyrmion stands for digital "0". In such a case, it is expected that information can be computed based on the manipulation of skyrmions and antiskyrmions. Indeed, there should be a detection scheme for the system containing both skyrmions and antiskyrmion. An effective detection method should be sensitive to the variation of in-plane spin textures of skyrmions, such as the electric Hall measurement (see Sec. 3.2). Namely, the skyrmions and antiskyrmions will be differentiated by the sign of the THE since the THE is proportional to the skyrmion number.

As mentioned in Sec. 1.1, the skyrmionium [24] has a topological charge of $Q = 0$. In 2013, the skyrmionium structure was experimentally created by a laser and observed in real space by Finazzi et al. [209]. Recently, Zhang et al. [323] also realized the real-space observation of a skyrmionium in ferromagnetic thin films coupled to a magnetic topological insulator [see Fig. 14(b)]. As the skyrmionium has a zero topological charge, it is immune from the SkHE and can be used to reliably delivery information in narrow and long nanotracks. Recently theoretical and simulation works have demonstrated that both spin currents and spin waves are able to drive skyrmioniums into motion in narrow nanotracks [58, 71, 72, 263].

The skyrmion structure with a topological charge of $|Q|$ = +2 is referred to as the biskyrmion. Theoretically, the skyrmion with $Q_v$ = +2 can be excited from a skyrmion with $Q_v$ = +1 by external stimuli, such as the STT [325]. The biskyrmion with $Q_v$ = +2 can also be formed by merging two skyrmions with $Q_v$ = +1, as theoretically demonstrated in frustrated magnets [302]. Recent years, some experiments revealed the rich static and dynamic properties of biskyrmions in different type of magnetic materials [326-330]. In 2014, Yu et al. [330] observed biskyrmions in a layered manganite by using the LTEM [see Fig. 14(c)], and realized the current-driven motion of biskyrmions. However, Loudon et al. pointed out in a recent report that the images of biskyrmions observed by LTEM can be explained as type-II magnetic bubbles viewed at an angle to their axes [331]. Note that the type-II magnetic bubble has a topological charge of zero and thus is a topologically trivial object. So, future works on biskyrmions could focus on their topology-dependent dynamics, which should be different from that of type-II bubbles.

The above mentioned antiskyrmion, skyrmionium, and biskyrmion are possible solutions in easy-axis magnets. In easy-plane magnets (i.e., magnets in which magnetization favor in-plane configuration), there is also a counterpart of skyrmions, i.e. the so-called bimeron [140]. The bimeron consists of a meron and an antimeron, which forms an asymmetric spin texture carrying a topological charge of $|Q|$ = 1. For this reason, the bimeron is also referred to as a meron-antimeron pair [84, 332, 333] or an asymmetric skyrmion in easy-plane magnets [334, 335]. The meron and antimeron were originated in classical field theory [336], of which the concept was then studied in different systems. For example, the bimeron was theoretically studied in quantum Hall systems in 2010 [337, 338], where a skyrmion in the bilayer system with imbalanced the electron density may be deformed into a bimeron when the parallel magnetic field penetrates between the two layers.

As a counterpart of the skyrmions in easy-axis magnets (i.e., magnets where magnetization favor out-of-plane configuration), the bimeron is a localized and compact quasi-particle like spin textures in easy-plane magnets. Recently theoretical works have suggested that an isolated bimeron in either easy-plane ferromagnets or easy-plane antiferromagnets can be driven into motion by STTs [334, 335, 339]. Similar to the transformation from strip magnetic domains to magnetic skyrmions or bubbles via the "pinching mechanism" [32, 34, 116, 184, 340], the mutual conversion between a skyrmion and a bimeron is also an important issue, which is essential for future spintronic circuits based on different topological spin textures. In 2014, Zhang et al. numerically demonstrated the possibility that a single isolated skyrmion in an easy-axis magnet can be transformed to a single isolated bimeron in an easy-plane magnet through a bridge-like narrow nanotrack junction connecting the easy-axis and easy-plane magnets [140]. In 2018, as shown in Fig. 14(d), the magnetic field-driven transformation between meron-antimeron pair lattice and skyrmion lattice was experimentally realized in a chiral magnet by Yu et al. [341, 342].

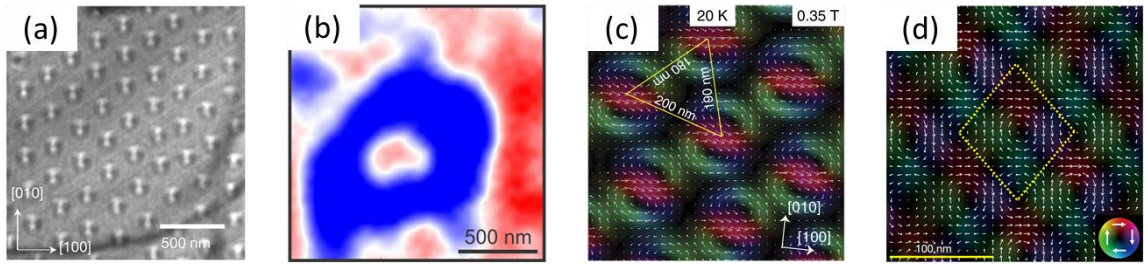

Figure 14. Other topological spin textures similar to skyrmions. (a) Under-focused LTEM images of antiskyrmions taken at fields applied along [001] of 0.29 T for antiskyrmions in $Mn_{1.4}Pt_{0.9}Pd_{0.1}Sn$. Reprinted with permission from [88]. (b) XPEEM image of an isolated skyrmionium obtained at the Fe L3 edge of the NiFe top layer at a temperature of 44 K in zero applied magnetic field. Reprinted with permission from [323]. (c) The spin texture of biskyrmion lattice in a bilayered manganese oxide $La_{2-2x}Sr_{1+2x}Mn_2O_7$ with $x = 0.315$. Reprinted with permission from [330]. (d) Real-space magnetization textures of a square lattice of merons and antimerons in a thin plate of the chiral-lattice magnet $Co_8Zn_9Mn_3$. Reprinted with permission from [341].

## 5.3. Summary

In this review, we have mainly focused on key device application-relevant findings and advances that have been made in the emerging field of *skyrmion-electronics* since the first experimental identification of skyrmions in 2009, which might indeed enable practical applications in the future. We have reviewed the writing and deleting of skyrmions by using different methods, including magnetic field, electric field, electric current, and laser. The imaging and electrical reading of skyrmions by different state-of-art experimental techniques are also discussed. Moreover, as the processing of skyrmions is fundamental to any skyrmion-based functional device applications, relevant findings have been reviewed with a focus on the implementation of skyrmion-based racetrack memory, logic computing devices and other emerging devices. The *skyrmion-electronics* is a rapidly growing field, in which a number of main topics have been expanded and many promising new topics have been introduced. Therefore, we have also introduced and put forward an overview on recent skyrmion research of different material systems as well as different skyrmion counterparts. We envision that *skyrmion-electronics* will continue to be an active and intriguing field of study, where more theoretical and experimental findings on the manipulation of skyrmions and their counterparts will emerge to support the development of practical applications and products for industry.

Finally, we note that no review articles can cover this topic exhaustively, and therefore, here we refer to other published reviews where some of the aspects of chiral magnetic skyrmions are discussed in more detail [2, 3, 5-7, 94, 98, 100, 157, 163, 295, 343-345].

## 6. Acknowledgements

X.Z. acknowledges the support by the Presidential Postdoctoral Fellowship of The Chinese University of Hong Kong, Shenzhen (CUHKSZ). Y.Z. acknowledges the support by the President's Fund of CUHKSZ, Longgang Key Laboratory of Applied Spintronics, National Natural Science Foundation of China (Grant Nos. 11974298 and 61961136006), Shenzhen Fundamental Research Fund (Grant No. JCYJ20170410171958839), and Shenzhen Peacock Group Plan (Grant No. KQTD20180413181702403). K.M.S. and T.-E.P. acknowledge the support by KIST Institutional Program (2E29410) and the support from the National Research Council of Science and Technology (NST) (Grant No. CAP-16-01-KIST) by the Korean Government (MSIP). M.E. acknowledges the support by the Grants-in-Aid for Scientific Research from JSPS KAKENHI (Grant Nos. JP18H03676, JP17K05490, and JP15H05854) and also the support by CREST, JST (Grant Nos. JPMJCR16F1 and JPMJCR1874). W.Z. acknowledges the support by the National Natural Science Foundation of China (Grant Nos. 61871008, 61571023, 61627813), and the National Key Technology Program of China (Grant No. 2017ZX01032101). G.Z. acknowledges the support by the National Natural Science Foundation of China (Grant Nos. 51771127, 51571126 and 51772004) of China, the Scientific Research Fund of Sichuan Provincial Education Department (Grant Nos. 18TD0010 and 16CZ0006). X.L. acknowledges the support by the Grants-in-Aid for Scientific Research from JSPS KAKENHI (Grant Nos. 17K19074, 26600041 and 22360122). S.W. acknowledges the support from IBM Research and management support from Guohan Hu and Daniel Worledge.